\documentclass[aps,prr,twocolumn,amsmath,amssymb,floatfix,
superscriptaddress]{revtex4-1}
\usepackage{blindtext}
\usepackage{epsfig}
\usepackage{color}
\usepackage{amsmath}
\usepackage{amsfonts}
\usepackage{amssymb}
\usepackage{graphicx}%
\usepackage{url}
\usepackage[breaklinks]{hyperref}
\usepackage[english]{babel}
\usepackage{esvect}
\usepackage[normalem]{ulem}
\usepackage[inline]{enumitem}
\usepackage{xfrac}
\usepackage{dsfont}
\usepackage{bm}

\hypersetup{colorlinks,citecolor=blue,filecolor=blue,linkcolor=blue,urlcolor=blue}

\numberwithin{equation}{section}
\renewcommand{\theequation}{\arabic{section}.\arabic{equation}}

\bibpunct{[}{]}{,}{n}{}{}

\begin{document}

\title{Valence bond fluctuations in the Kitaev spin model}

\author{Fan Yang}
\affiliation{CPHT, CNRS, Institut Polytechnique de Paris, Route de Saclay, 91128 Palaiseau, France}
\author{Kirill Plekhanov}
\affiliation{CPHT, CNRS, Institut Polytechnique de Paris, Route de Saclay, 91128 Palaiseau, France}
\affiliation{LPTMS, Universit\'{e} Paris-Sud, CNRS, Universit\'{e} Paris-Saclay, 91405 Orsay C\'{e}dex, France}
\affiliation{Department of Physics, University of Basel, Klingelbergstrasse 82, CH-4056 Basel, Switzerland}
\author{Karyn Le Hur}
\affiliation{CPHT, CNRS, Institut Polytechnique de Paris, Route de Saclay, 91128 Palaiseau, France}

\begin{abstract}
  We introduce valence bond fluctuations, or bipartite fluctuations
  associated to bond-bond correlation functions, to characterize
  quantum spin liquids and the entanglement properties of them. Using
  analytical and numerical approaches, we find an identical scaling
  law between valence bond fluctuations and entanglement entropy in
  the two-dimensional Kitaev spin model and in one-dimensional chain
  analogues. We also show how these valence bond fluctuations can
  locate, via the linear scaling prefactor, the quantum phase transitions between the three gapped and the
  gapless Majorana semi-metal phases in the honeycomb model. We then
  study the effect of a uniform magnetic field along the $[111]$
  direction opening a gap in the intermediate phase which becomes
  topological. We still obtain a robust signal to characterize the
  transitions towards the three gapped phases. The area law  behavior of such bipartite fluctuations in two dimensions is also
  distinguishable from the one in the N\' eel magnetic state that follows a volume square growth.
  
\end{abstract}

\date{\today}

\maketitle

\section{Introduction}

The quest of quantum spin liquids in the Mott
regime~\cite{haldane19831, haldane19832, anderson1987, affleck1987,
  balents2010, pollmann2012} has been a great challenge these last
decades in relation with the discovery of {quantum
  materials}~\cite{renard1988, alloul1989, david1989, hagiwara1990,
  coldea2001, shimizu2003, mendels2010, mendels2016, takagi2019}.
Quantum spin liquids show interesting topological and entanglement
properties~\cite{kalmeyer1987, kivelson1987, wen1989, read1991} which
can be used for applications in quantum
information~\cite{kitaev2009}. The Kitaev spin model on the honeycomb
lattice~\cite{kitaev2006} represents an important class of models,
since it can be solved exactly in a Majorana fermion representation
and demonstrates the significance of $\mathbb{Z}_2$ gauge fields on
the low-energy properties. The model shows three gapped spin liquid
phases carrying Abelian anyon excitations and an intermediate gapless
phase which can be identified as a semi-metal of Majorana
fermions. Applying a magnetic field along the three spatial
directions~\cite{kitaev2006}, referring to a field in the $[111]$
direction, induces a gap in the intermediate phase, then producing a
topological $\mathbb{Z}_2$ phase supporting non-Abelian anyonic excitations~\cite{MooreRead} 
and closely related to a $p_x+ip_y$ superconductor~\cite{ReadGreen} with {chiral edge modes}. 
It is important to emphasize that static spin-spin correlation functions are exactly
zero {beyond nearest neighbors} in this model~\cite{baskaran2007}. 
Theoretical efforts have been performed to compute dynamical correlation 
functions~\cite{kitaev2011,knolle2014, song2016, kourtis, pollmann} as well as the entanglement
entropy~\cite{yao2010,pollmann}.  By bipartitioning a system
spatially, the entanglement entropy measures how entangled the two
subsystems are~\cite{eisert2010}. Related to these theoretical
developments, quasi-two-dimensional quantum materials have been
{synthesized}~\cite{Kee,Trebst,Takagi}, with recent measurements from
neutron~\cite{neutron} and Raman~\cite{Nasu} scatterings, nuclear
magnetic resonance~\cite{janvsa2018} and thermal transport~\cite{nasu2017, kasahara20181, kasahara20182}. 
One and two-dimensional Kitaev spin liquids, could also be engineered in ultra-cold atoms~\cite{Duan}
and quantum circuits ~\cite{circuit,Oreg}.

In this {Article}, we propose valence bond fluctuations as a probe of
entanglement properties in the ground state of the
Kitaev spin model.

A valence bond (VB)~\cite{anderson1987} here corresponds to the spin-spin
pairing between two nearest neighbor electrons.  
Our first insight comes from the system of SU(2)-symmetric quantum spins with resonating valence
bonds (RVB) where we find the bond fluctuations can be related to valence bond entropy of Einstein-Podolsky-Rosen 
(EPR) pairs or Bell pairs~\cite{alet2007}. 
Extending to the Kitaev spin liquids, in the three gapped
phases, the valence bonds between nearest neighbors form a crystalline
or dimer order~\cite{kitaev2006}. Approaching the transition(s) to the
gapless intermediate phase, these bonds now resonate giving rise to
gapless critical fluctuations, which in principle encode information on 
quantum phase transitions and entanglement properties.
Our calculations indeed reveal an
identical scaling between valence bond
fluctuations and entanglement entropy in one-dimensional chain
 and two-dimensional honeycomb lattice, and we check our mathematical
findings with numerical calculations, {\it e.g.} through the Density
Matrix Renormalization Group (DMRG). In one dimension, the gapless
phase is reduced to a quantum critical point~\cite{feng2007}, which
then develops {into a plane} for ladder systems~\cite{PRBlong}. In two
dimensions, in the absence of a magnetic field, the long-range valence bond
correlations in space~\cite{shuo2008} share a similar scaling as the
dynamical spin structure factor~\cite{kitaev2011}. 
We include also the effects
of a uniform magnetic field in the perturbative regime, and discuss relevant
consequences from the excitations of flux pairs~\cite{song2016} and from the formation of U(1)
gapless spin liquids once three Ising couplings become anti-ferromagnetic (AFM)~\cite{hickey2019}.
In the end, to make a closer link with quantum materials, 
we give a comparison of valence bond fluctuations in the N{\'e}el state favored by strong AFM Heisenberg
exchanges. 
%-----------------------------------------------------------------------------------------
\section{Review of Fluctuations as An Entanglement Probe }

First, we begin with a brief review on the relation between bipartite fluctuations and entanglement entropy in many-body Hamiltonians characterized by different symmetries~\cite{song2012, alex2014, loic2017,pollmann2014}.  
In Sec.~\ref{sec:gen}, we define the general fluctuations on a bipartite lattice, which from the information theory,  
provide a lower bound for the mutual information related to entropy. We then remind in Sec.~\ref{sec:example},  an exact expression for the entropy as a series expansion of even cumulants (with particle number or spin fluctuations 
the leading order) for the U(1) charge conserved systems \cite{song2012}  and 
an inequality between the two quantities emerging among the SU(2) quantum spins  described by
resonating valence bonds. 

Generalizing these works to Kitaev spin liquids coupled to a gapped
 $\mathbb{Z}_2$ gauge field represents our central motivation in this work. The difficulty lies in finding the right observable encoding the long-range correlation
of matter Majorana fermions in the gapless phase, hence the entanglement properties. 
Fortunately, by analogy to the RVB states in the three gapped phases, in Sec.~\ref{sec:lowerbound} we verify that the valence bond fluctuations represent
non-vanishing lower bound for the entropy. 

\subsection{Generalities}
\label{sec:gen}
We decompose a generic quantum system into two parts $A \cup B$.
For the subsystem $A$, the entanglement is measured by the von Neumann entropy~\cite{eisert2010} 
  \begin{gather}
    \mathcal{S}_A = -\text{Tr} {\rho}_A \ln {\rho}_A,
  \end{gather}
where ${\rho}_A = \text{Tr}_B \rho$ represents the reduced density matrix of sub-system $A$. Once given two density matrices
$\rho$ and $\rho'$, the distance between two states can be probed by the relative entropy 
  \begin{gather}
    \mathcal{S}(\rho,\rho') = \text{Tr} \left[ \rho(\log \rho - \log \rho')\right], \label{eq:re1}
  \end{gather}
 with a norm bound~\cite{ohya2004}
    \begin{gather}
     \mathcal{S}(\rho,\rho') \ge \frac{1}{2} \left|| \rho - \rho'|\right|^2. \label{eq:re2}
   \end{gather}
 Here the norm stands for $\left|| \rho |\right| = \text{Tr} \sqrt{\rho^\dagger \rho}$ and we have assumed that $\hbox{Tr}\rho=\hbox{Tr}\rho'=1$. Making an analogy with vectors, one may also write $S(\rho,\rho')=S(\rho\parallel\rho')$. For instance, for diagonal (density) matrices, each eigenvalue may refer to a coordinate along one direction.  On the other hand, to evaluate the fluctuations, we introduce two measurements
\begin{gather}
  \mathcal{F}_A = \left< \left( \sum_{i \in A} Q_i \right)^2\right>_c, \label{eq:fa} \\
   \mathcal{F}_{AB} = \left| \sum_{i \in A} \sum_{j \in B} \left< Q_i Q_j \right>_c \right|. 
\label{eq:fab}
\end{gather}
Here, $Q$ is a chosen operator for targeted systems: charge, particle number, one spin or two spins on a valence bond and $\left< Q_i Q_j \right>_c = \langle Q_i Q_j \rangle - \langle Q_i \rangle \langle Q_j \rangle$ denotes the reduced 
correlation function. 
It is easy to notice while $\mathcal{F}_A$ measures the fluctuations in subsystem $A$,
$\mathcal{F}_{AB}$  covers the correlations between $A$ and $B$.
There is an equality between the two quantities:
 \begin{gather}
   \mathcal{F}_{AB} = \frac{1}{2} \left|  \mathcal{F}_A +
    \mathcal{F}_B - \mathcal{F}_{A \cup B}  \right|. \label{eq:ffab}
 \end{gather}
 
 An important finding so far established is to relate $\mathcal{F}_{AB}$ with $\mathcal{S}_A$ by mutual information~\cite{nishioka2018}
   \begin{gather}
     I(A, B) = \mathcal{S}_A + \mathcal{S}_B - \mathcal{S}_{A \cup B}.
   \end{gather}
From the definition of relative entropy (\ref{eq:re1}), the mutual information has an alternative expression
   \begin{gather}
    I(A, B) = \mathcal{S}(\rho_{A \cup B} , \rho_A \otimes \rho_B).
   \end{gather}
Choosing any operator of the matrix form $Q = Q_A \otimes Q_B$ with $Q_A$ the bounded operator in region $A$ and applying the Schwarz inequality $\left|| \rho |\right| \ge \text{Tr} (\rho Q) / \left||Q|\right|$ to the norm bound (\ref{eq:re2}), 
one obtains~\cite{wolf2008}
  \begin{gather}
    I(A,B) \ge \frac{\left( \langle Q_A Q_B \rangle - \langle Q_A \rangle \langle Q_B \rangle \right)^2}{2\left||Q_A|\right|^2 \left||Q_B|\right|^2}. 
  \end{gather}
The numerator recovers $\mathcal{F}_{AB}$. Correspondingly, for $\mathcal{F}_{AB} \ne 0$, we arrive at
  \begin{gather}
    \frac{I(A, B)}{ \mathcal{F}_{AB}} \ge \text{cst}. \label{eq:lb}
  \end{gather}
Although the lower bound between the bipartite fluctuations and the mutual information is universal, 
it remains ambiguous what is the form of operator $Q$ one should choose for a given many-body system such that the fluctuations measured are non-vanishing. 
A second inquiry would be: under which circumstances we could reach the equality of  (\ref{eq:lb}) such that the fluctuations share the same scaling as the original entropy.

%----------------------------------------------------------------------------------------
\subsection{Exact relations and inequalities}
\label{sec:example}
In this subsection, we give two known examples where one can relate entanglement entropy  directly to charge or spin fluctuations. We further show in Sec.~\ref{sec:epr}, for the SU(2)-symmetric RVB state, that the bond correlator is also a good option for operator $Q$.

\subsubsection{Noninteracting fermions with conserved U(1) charge}

Let us consider a system of non-interacting fermions with conserved total charge, or more precisely total number of particles: $\hat{N}_{\text{tot}} | \psi \rangle  = N_{\text{tot}} | \psi \rangle$.
At zero temperature, the ground state $| \psi \rangle$ of the total
system becomes pure.
Basic properties of 
the entanglement entropy then follow
 \begin{itemize}
   \item \textsl{symmetric}: $\mathcal{S}_A = \mathcal{S}_B$;
   \item \textsl{subadditive}: $\mathcal{S}_A + \mathcal{S}_B \ge \mathcal{S}_{A\cup B} = 0$.
 \end{itemize}
Without any calculation, one can already see the similarities between charge fluctuations and the entropy. If we take $\hat{Q}_A = \hat{N}_A$ in $\mathcal{F}_A$ (\ref{eq:fa}), where $\hat{N}_A$ represents the number of particles in sub-region $A$, as a result  of total charge conservation $\hat{N}_A - N_A = -(\hat{N}_B - N_B)$, the fluctuations also inherit the symmetric and sub-additive characteristics
   \begin{gather}
     \mathcal{F}_A = \mathcal{F}_B, \qquad
     \mathcal{F}_A + \mathcal{F}_B = 2\mathcal{F}_A \ge 0.
   \end{gather}
   
In fact, one can relate the two quantities more rigorously by the cumulant expansion of entropy~\cite{song2012}
  \begin{gather}
    \mathcal{S}_A = \lim_{K \to \infty} \sum_{n=1}^{ \lfloor (K+1)/2 \rfloor} \alpha_{2n} (K) C_{2n}. \label{eq:sc}
  \end{gather}
Here, the  coefficients  are all positive and  related to the unsigned Stirling number of the first kind: $\alpha_{n}(K) = 2\sum_{k=n-1}^K S_1(k,n-1)/(k! k )$. The cumulants $C_n$ are given by the generating function $\chi(\lambda) = \langle e^{i\lambda \hat{N}_A} \rangle$ 
according to
  \begin{gather}
   C_n  = \left. (-i \partial_\lambda)^n \ln \chi(\lambda) \right|_{\lambda = 0}.
  \end{gather}
By definition, one verifies $\mathcal{F}_A = C_2$. For a Gaussian process, one can truncate the
serie with $K=1$, but for non-Gaussian models one needs to check carefully the convergence of the series  with the appropriate number of cumulants \cite{song2012}. As a comparison, if we consider a Bell pair or an EPR pair,
then this generally requires around 10 cumulants to reproduce the $\ln 2$ entropy.  Although the equivalence of entropy and a complete set of even cumulants 
$(\ref{eq:sc})$ is unique to the systems with a mapping to non-interacting fermions, the general relation between entropy and
fluctuations (\ref{eq:sf}) can be further extended to the interacting one-dimensional (1D) critical systems that conserve total charge, and can be described by a Gaussian model through conformal field theory (CFT) or bosonization. 
In those cases, $\mathcal{S}_A$ can also be truncated by a $K=1$ upper-bound. Thus one gets
 \begin{gather}
   \frac{\mathcal{S}_A}{\mathcal{F}_A}  \simeq \text{cst}.  \label{eq:sf}
 \end{gather}
The constant proves to encode rich information, for instance~\cite{song2012}

  \begin{gather}
    \text{cst} \cdot \frac{3}{\pi^2} = \begin{cases}
                       K , & \text{Luttinger liquids}; \\
                       c/ g, &  \text{U(1) CFTs},
                      \end{cases}
  \end{gather}
where similarly to Refs. \cite{song2012,song2011}, we also introduce the letter $K$ for the Luttinger parameter and $c$ represents the central charge in conformal field theory (CFT). Parameter $g = \pi v \kappa$ consists of  the velocity $v$ and compressibility $\kappa = \partial n /\partial \mu$.

%------------------------------------------------------------------------------------------
\subsubsection{SU(2) quantum spins with EPR pairs}
\label{sec:epr}

Intuitively, one may wonder what will happen when the system breaks U(1) charge conservation and when the system becomes higher dimensional, such as two-dimensional.  Next, we give an example of the SU(2)-symmetric valence bond state~\cite{song2012}. Indeed, a direct correspondence in terms of inequality similar to relation $(\ref{eq:sf})$, subsists if we replace $\mathcal{F}_{A}$ with $\mathcal{F}_{AB}$, by analogy to mutual information (\ref{eq:lb}). Furthermore, we would like to extend the result of Ref.~\cite{song2012} and show how the two-spin fluctuations and valence bond fluctuations capture different features of the entanglement entropy. Here, by ``two-spin fluctuations'', we refer to fluctuations associated to spin-spin correlation functions. 
From Eq.~(\ref{eq:fab}), two-spin (TS) fluctuations and valence bond fluctuations between two subregions read:
  \begin{eqnarray}
    \mathcal{F}^{\text{TS}} &=& \mathcal{F}_{AB} (Q_{i,\alpha} = \sigma^z_{i,\alpha}), \notag \\
     \mathcal{F}^{\text{VB}} &=& \mathcal{F}_{AB} (Q_i = \sigma^z_{i,1}\sigma^z_{i,2}). \label{eq:fsu2}
  \end{eqnarray}
In this work, our aim is to show that the valence bond fluctuations are essential since they provide relevant information both for SU(2) and $\mathbb{Z}_2$ quantum spin liquids. 

We consider the two-dimensional Heisenberg antiferromagnetic (HAF) model where arise two competing phases: 
a N{\'e}el state and a gapped VB state. In either configuration, the valence-bond entropy has proven to exhibit distinct behaviors~\cite{alet2007}:
  \begin{gather}
    \mathcal{S}^{\text{VB}}_{\text{2D HAF}} \sim 
      \begin{cases}
        ax \ln x + bx  & (\text{N\'eel}) \\
        b'x & (\text{VB})
      \end{cases}. \label{eq:vb}
  \end{gather}
Here $x$ denotes the length of the boundary between two subsystems. This measure can also accurately detect quantum phase transitions between N\' eel
and RVB spin phases in quasi-one dimensional ladder systems~\cite{stephan}. For the moment, we focus on the fluctuations of the gapped VB state and will address the comparison with a N{\'e}el state later in Sec.~\ref{sec:neel}.

Suppose our system comprises $N$ sites on even and odd sublattices.  
Dimer coverings between different sublattices sharing the form
  \begin{gather}
    |\frown\  \rangle_{\overline{(i,1)(j,2)}} = \frac{1}{\sqrt{2}} ( | \uparrow_{i,1} \rangle | \downarrow_{j,2} \rangle - | \downarrow_{i,1} \rangle | \uparrow_{j,2} \rangle),
   \end{gather}
 minimize the energy from the antiferromagnetic Heisenberg interactions.   
The subscript $(i,\alpha)$ describes the site on $i$-th unit cell of the $\{ \alpha  = 1,2 \}$ sublattice.
A singlet state $|\Phi_0\rangle$ can then be represented as a complex superposition of all possible dimer or pairing configurations
  \begin{gather}
     |\Phi_0\rangle  = \sum_p \lambda_p |\varphi_p\rangle, \notag \\
      | \varphi_p \rangle  = \prod_{\overline{(i,1)(j,2)} \in p}  |\frown\  \rangle_{\overline{(i,1)(j,2)}} . \label{eq:sing}
   \end{gather}
 For a given global pairing distribution $p$, the product state goes over all local internal dimers $\overline{(i,1)(j,2)}$. 
 
 The corresponding VB entanglement entropy (\ref{eq:vb}) is defined as~\cite{alet2007}
  \begin{gather}
    \mathcal{S}^{\text{VB}} (\Phi_0) = \frac{\sum_p \lambda_p \mathcal{S}^{\text{VB}} (\varphi_p)}{\sum_p \lambda_p}. 
  \end{gather}
Formally, acting on the singlet state $| \Phi_0 \rangle$, the fluctuations and the VB entropy can be obtained from the decomposition (\ref{eq:sing})
  \begin{align}
    \mathcal{F}^{\text{TS/VB}} (\Phi_0) &= \sum_p \sum_{p'} \lambda_p^* \lambda_{p'} \mathcal{F}^{\text{TS/VB}} (\varphi_p, \varphi_{p'}), \notag \\ 
    \mathcal{S}^{\text{VB}} (\Phi_0) &= (\ln 2) \cdot \frac{\sum_p \lambda_p  \mathcal{F}^{\text{TS}} (\varphi_p, \varphi_{p}) }{\sum_p \lambda_p}. \label{eq:fe}
   \end{align}
 Eq.~(\ref{eq:fe}) indicates in general there is no simple correspondence between $ \mathcal{F}^{\text{TS/VB}}$ and $\mathcal{S}^{\text{VB}}$. Yet, we may try to simplify 
  ($\ref{eq:fe}$) as
  \begin{gather}
    \mathcal{F}^{\text{TS/VB}} (\Phi_0) := \sum_p  |\lambda_p|^2 \mathcal{F}^{\text{TS/VB}} (\varphi_p, \varphi_{p}). \label{eq:se}
  \end{gather}
For $\mathcal{F}^{\text{TS}}$, the relation (\ref{eq:se}) is exact owing to the sublattice symmetry of two-spin correlation functions. For $\mathcal{F}^{\text{VB}}$, it is a redefinition in the sense one counts the bond fluctuations inside each pairing pattern with probability $|\lambda_p|^2$ and at the same time, ignores the contributions from the overlaps of different pairing patterns. This  redefinition  is crucial for $\mathcal{F}^{\text{VB}}$ to resemble the behavior of the VB entanglement entropy.

We present in Appendix~\ref{app:su2} a detailed analysis of the fluctuations in SU(2)-symmetric quantum spin systems.

On one hand, if the gapped VB state is composed of $N$-site singlets carrying equal weights ($\lambda_p = \lambda$), we have the following inequality reminiscent of the lower
bound for mutual information (\ref{eq:lb})
      \begin{gather}
        \mathcal{S}^{\text{VB}} \propto \ln 2 \cdot n, \notag \\
          \frac{\mathcal{S}^{\text{VB}}}{\mathcal{F}^{\text{TS}} + \mathcal{F}^{\text{VB}}} \ge \ln 2.
      \end{gather}
Here $n$ denotes the number of singlets that the boundary crosses. Both relations above take the equality $``="$ if the maximum resonating range $N \le 4$.  As $N \to \infty$, the system approaches the gapless critical point.

On the other hand,  as soon as the singlet bonds decay exponentially  with distance ($\lambda_p \sim e^{-r/\xi}$),
  \begin{gather}
    \mathcal{F}^{\text{TS/VB}}_{\text{2D HAF}}  = b^{\text{TS/VB}}  \cdot x + \mathcal{O}(x).  \label{eq:ham}
   \end{gather}
A similar area law scaling is revealed in two types of fluctuations alongside the VB entanglement entropy (\ref{eq:vb}).

%-----------------------------------------------------------------------------------------------------------------------------------------------------
\subsection{Generalization to Kitaev $\mathbb{Z}_2$ spin liquids}
\label{sec:lowerbound}

Valence bond states in SU(2) spin systems and Kitaev $\mathbb{Z}_2$ spin liquids may be distinguishable from the form of correlation functions.  For the former, the two-spin correlator follows an exponential decay in the gapped phase; for the latter, however, the static correlation between two spins becomes exactly zero beyond nearest neighbors~\cite{baskaran2007}.

In fact, the Kitaev honeycomb model is solved in the Majorana representation with one spin operator mapped onto 
the product of one matter and one gauge Majorana fermions~\cite{kitaev2006}.
Once acting on the ground state embedded with a static $\mathbb{Z}_2$ gauge field, the gauge Majorana fermion creates a pair of fluxes in two adjacent
hexagons. It renders two-spin fluctuations irrelevant, if given an arbitrary boundary (not assigned on the same Ising links)
  \begin{gather}
     \mathcal{F}^{\text{TS}}_{\text{Kitaev}, \mathbb{Z}_2} = 0. \label{eq:tsk}
  \end{gather}
One should resort to the bond-bond operator~\cite{feng2007}. Since the excitation of a flux pair is annihilated simultaneously by the other spin on the same bond,
valence bond fluctuations always give a relevant lower bound regardless of the boundary position
  \begin{gather}
     \frac{\mathcal{S}_{F}}{\mathcal{F}^{\text{VB}}_{\text{Kitaev}, \mathbb{Z}_2}} \ge \text{cst}. \label{eq:vfb}
  \end{gather}
  In the equation above, we have rescaled the mutual information by Fermi entropy according to the area law of entanglement entropy~\cite{yao2010}.

Another observation comes consistently from the SU(2)-invariant Kitaev spin liquids~\cite{yao2011}: there, the two-spin operator can be expressed solely in terms of matter Majorana fermions (preserving the gauge structure) and its correlation becomes non-vanishing. Similar to the Heisenberg anti-ferromagnet  (\ref{eq:ham}), the spin and bond fluctuations obey a linear growth in the gapped region
  \begin{gather}
    \mathcal{F}^{\text{TS/VB}}_{\text{Kitaev, SU(2)}}  = b^{\text{TS/VB}}  \cdot x + \mathcal{O}(x). 
   \end{gather}

We then conclude that both two-spin and valence bond fluctuations are appropriate as relevant probes of the entanglement entropy for the SU(2) spin systems, whereas for the Kitaev $\mathbb{Z}_2$ spin liquids only the valence bond fluctuations play a substantial role. 

Below, we address valence bond fluctuations both for one-dimensional and two-dimensional Kitaev spin models. In Secs.~\ref{sec:chain} and \ref{sec:honeycomb}, we prove how valence bond fluctuations (\ref{eq:vfb}) develop the same scaling as the entanglement entropy both for the one-dimensional chain and the honeycomb lattice.

%--------------------------------------------------------------------------------------------------------------------------------------------------------
\section{Model on the chain} 
\label{sec:chain}

Here, we address the quantum chain or
wire model~\cite{feng2007,PRBlong}, shown in Fig.~\ref{fig:chain}
(a). The Hamiltonian takes the form
  \begin{gather}
    \mathcal{H} = \sum_{j=2m-1} J_1\sigma_j^x\sigma_{j+1}^x +
    J_2\sigma_{j+1}^y \sigma_{j+2}^y, \label{eq:hsc}
  \end{gather}
  The sum acts on odd sites only such that $ 1 \leq m\leq M$ is an
  integer with $M$ being the total number of unit cells. 

\subsection{Valence bond  correlator}

Applying the
  Jordan-Wigner transformation, one can then map quantum spins-1/2
  onto spinless fermionic operators with occupancy $1$ and $0$ at each
  site:
  $\sigma_j^+ = a_j^\dagger  \cdot {\prod_{i<j}(-\sigma_i^z)},
  \sigma_j^- = a_j  \cdot {\prod_{i<j}(-\sigma_i^z)}, \sigma_j^z =
  2a_j^\dagger a_j -1$. We further consider the following
  representation with two Majorana fermions per site:
    \begin{align}
       c_j &= i(a_j^{\dagger}-a_j), \quad d_j=a_j^{\dagger}+a_j, \quad  j=2m-1; \notag \\
       c_j &= a_j^{\dagger}+a_j,  \quad d_j=i(a_j^{\dagger}-a_j), \quad  j=2m. 
     \end{align}
   The key point is that in this basis all the $d$ Majorana fermions
  decouple from the chain and encode the double-degeneracy of the
  (spin) ground state on a given bond of nearest neighbors. A $d$
  Majorana fermion contributes to a {$(\ln 2)/2$} entropy by
  analogy to the two-channel Kondo
 model~\cite{affleck1991,2CKME}. Below, we address long-range quantum
  properties of the chain captured by the $c_j$ Majorana
  fermion Hamiltonian~\cite{PRBlong}
  \begin{equation}
  \mathcal{H} = -i \sum_{j = 2m - 1} (J_1 c_j c_{j + 1} - J_2 c_{j + 1}c_{j + 2}).
  \label{eq:hc0}
  \end{equation}
  
To calculate valence bond correlation functions, it is useful to introduce a complex bond fermion operator acting in the middle of two sites $(2m-1, 2m)$
  \begin{gather}
     \psi_{m} = \frac{1}{2}\left( c_{2m-1} + i c_{2m}\right),
  \end{gather}
thus forming a dual lattice
with site index $m$.
  In momentum space, choosing the basis
$\Psi^\dagger = ( \psi_k^\dagger, \psi_{-k} )$, the Hamiltonian
becomes
 \begin{gather}
    \mathcal{H} = \sum_k \Psi^\dagger \widetilde{M} \Psi, 
     \qquad
    \widetilde{M} =
    \begin{pmatrix}
      \xi_k & -  \Delta_k \\
      -\Delta_k^* & -\xi_k
    \end{pmatrix}.  \label{eq:hkc} 
  \end{gather}
  The matrix elements read 
    \begin{gather}
      \xi_k = \mathfrak{Re}{f(k)}, \quad \Delta_k = i \mathfrak{Im} f(k), \quad
      f(k) = -J_1-J_2e^{-2ikl}.
      \end{gather}
From now on, we set the  lattice spacing  $l$ of the original chain to
  unity  $1$.   
 Matrix $\widetilde{M}$ has two eigenvalues
    \begin{gather}
      (E_k^\pm)^2 = |\xi_k|^2 + |\Delta_k|^2, \label{eq:ev} \\
      E_k^{\pm} =      \pm \sqrt{J_1^2 + J_2^2 + 2J_1J_2 \cos(2k)}, 
    \end{gather}
    reflecting a gap in
  the spectrum if $J_1\neq J_2$. We check that the gap closes at
  {$k_F = \pi/2$ when $J_1 = J_2$} and that the  chain on
  the dual lattice (\ref{eq:hkc}) results in {a critical gapless theory of free fermions  with central charge
    $c=1$. We find its counterpart in the original spin basis through
  the observable ``valence bond correlator''. 
  
Using definitions in Sec. \ref{sec:gen}, we introduce the bond-bond correlation functions $I(i,j)=\langle Q_i Q_j\rangle_c$, with here: 
 \begin{gather}
     Q_j = \sigma_{j,1}^x \sigma_{j,2}^x = -ic_{j,1}c_{j,2}.
     \label{eq:bo}
   \end{gather}  
It is important to underline that in Fig. \ref{fig:chain} (a), we have chosen the strong bonds associated to the $J_1$ coupling, referring to the $x$ spin Pauli operator in $Q_j$.

As before, the site
  index $(j,\alpha)$ represents the $j$-th unit cell of the
  sublattice $\alpha = \{ 1, 2 \}$. In the dual lattice, $Q_j$
  relates to the density of bond fermions $\psi^{\dagger}_j\psi_j$.
  At the gapless point, we get from Wick's theorem
      \begin{gather} 
       I(i,j) = I(|i-j|) = \frac{1}{\pi^2}  \frac{1}{|i-j|^2 - 1/4}, \label{eq:ik1}
    \end{gather}
for $i \ne j$.
As a comparison, take the usual two-spin correlator
$\langle \sigma_i^z \sigma_j^z \rangle = (-1)^{i+j+1} \left<
  c_id_ic_jd_j\right>$.  Decoupled $d$ Majorana fermions lead to
$\langle c_id_j \rangle = \langle d_i d_j \rangle = 0$. Applying
Wick's theorem, $\langle \sigma_i^z \sigma_j^z \rangle$ vanishes in
all phases beyond nearest neighbors, and the same for $\langle \sigma_i^x \sigma_j^x \rangle$
and $\langle \sigma_i^y \sigma_j^y \rangle$ which involve
Jordan-Wigner strings formed by the pairings of $c$ and $d$ Majorana
fermions.  Once the distance of two sites goes beyond the nearest neighbour $|i - j| > 1$, 
one verifies
  \begin{gather}
    \langle \sigma_i^a \sigma_j^a \rangle = 0, \quad a = x,y,z.
  \end{gather}
Like the Kitaev honeycomb model, in its one-dimensional chain analogue, again the two-spin operator does not encode the long-range correlation of the gapless Majorana fermions.

Now deviating from the gapless point, from the bond-fermion
model and from the Ising symmetry of the spin chain, we predict that
the correlation length $\xi$ of the bond operator is proportional to the inverse of the gap
$\Delta = |J_2 - J_1|$,
  \begin{gather}
    \xi \propto \Delta^{-\nu}, \quad \nu=1. \label{eq:kci1}
 \end{gather}
The valence bond correlations share the behavior 
  \begin{gather}
    I(i,j) = \begin{cases}
                 c_1 | i-j |^{-2}, \quad &| i-j | \le \xi ; \\
                 c_2 e^{- {| i-j |}/{\xi}}, \quad &\text{otherwise}. 
              \end{cases} \label{eq:kci2}
  \end{gather}
We then perform numerical {calculations} based on DMRG which verify these predictions
with the associated critical exponent $\nu = 0.94\sim 1$ in the inset
of Fig.~\ref{fig:chain} (c).

\begin{figure}[t]
    \begin{center}
      \includegraphics[width=0.3\textwidth]{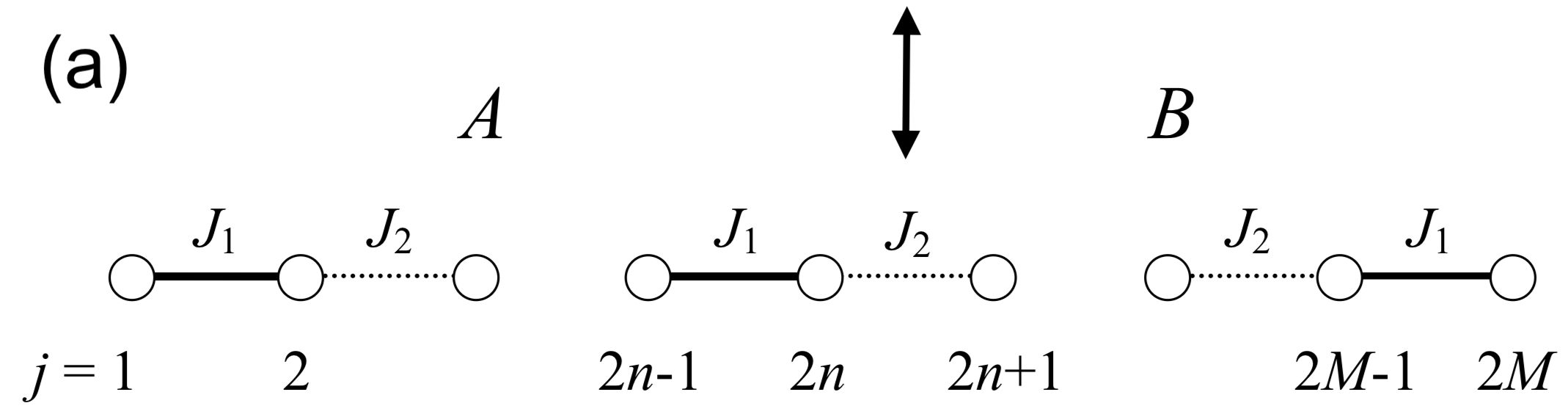} \\
       \vspace{0.3cm}
      \includegraphics[height=2.8cm]{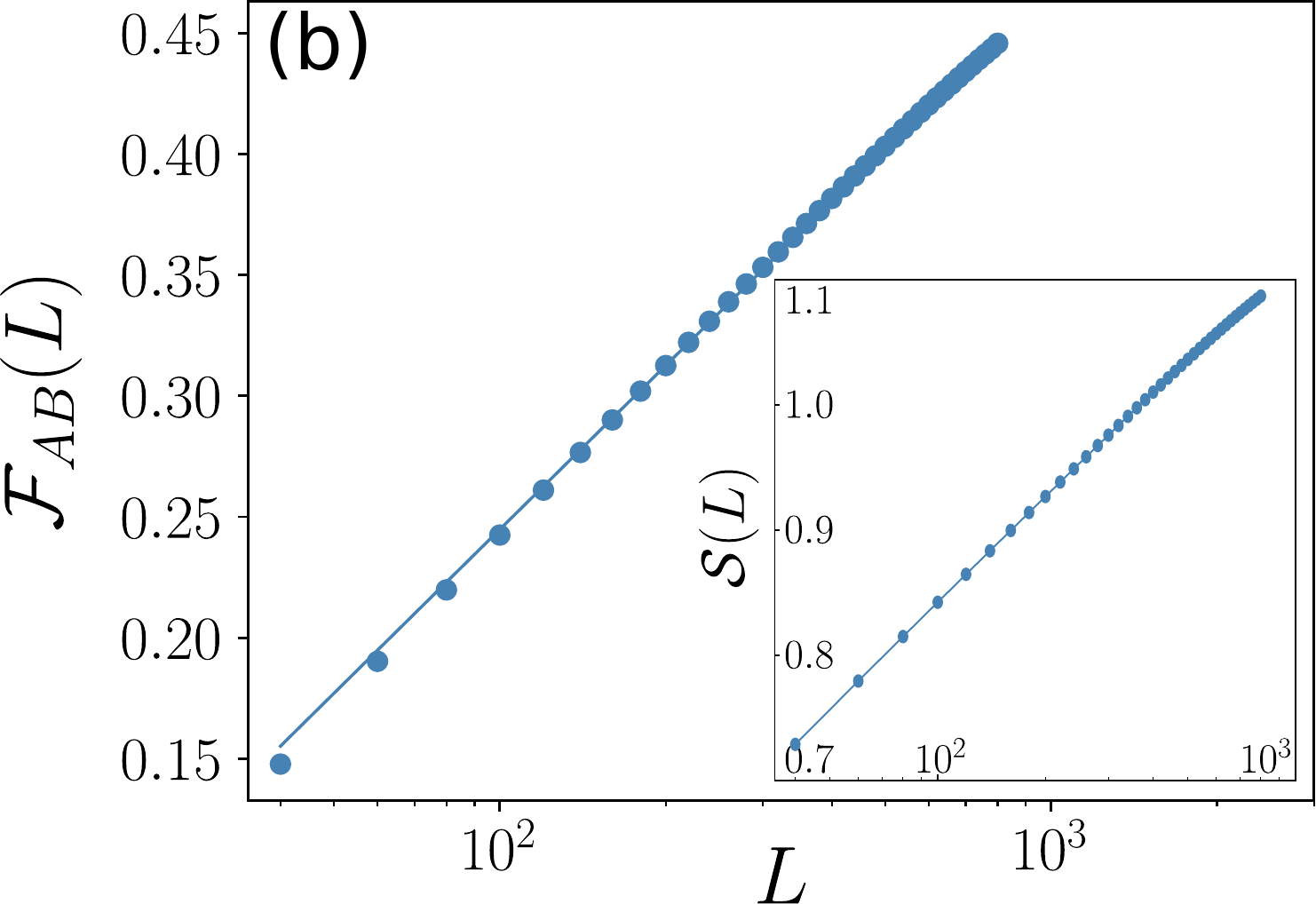}  \quad
      \includegraphics[height=2.82cm]{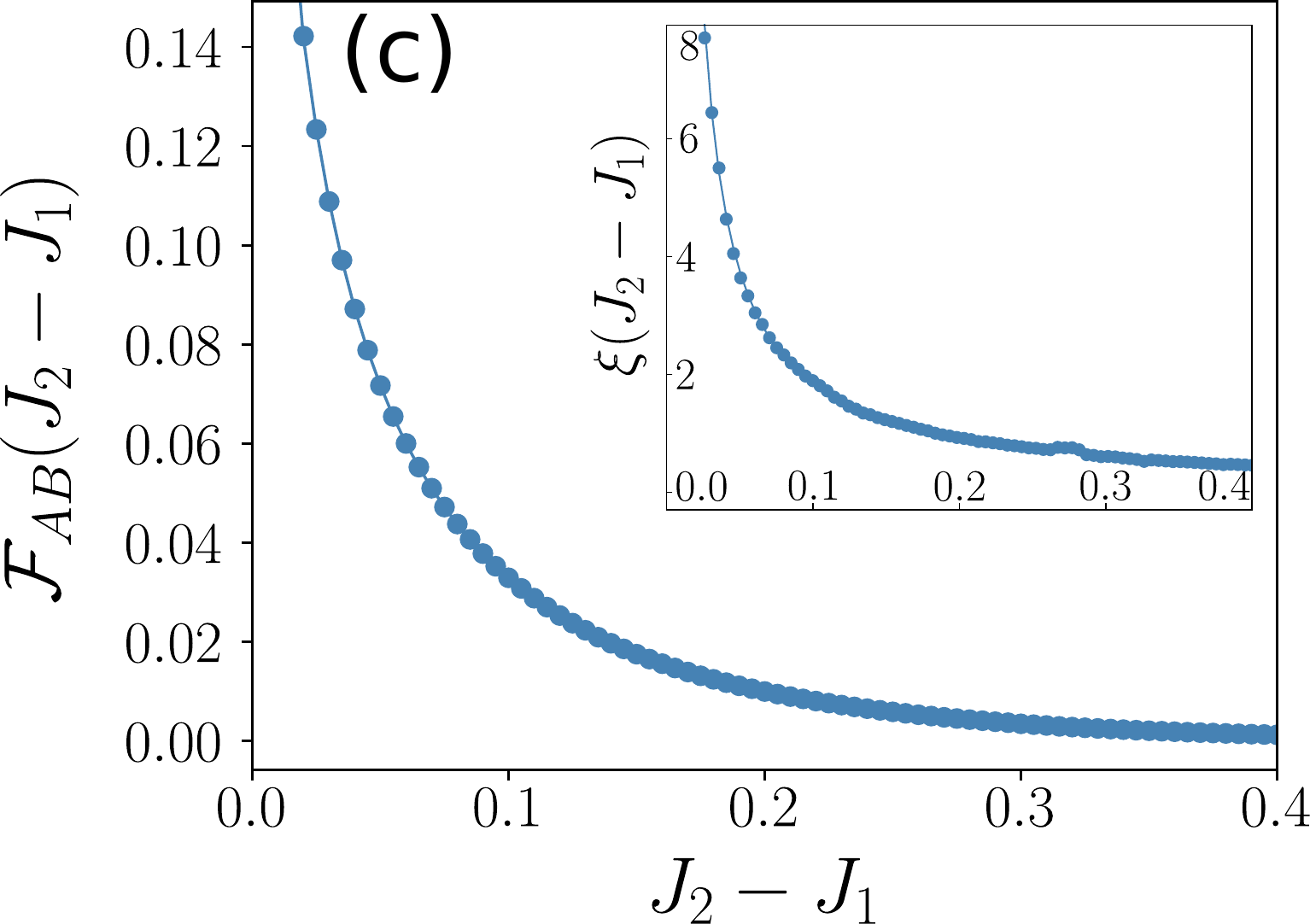}     
   \end{center}  
   \vskip -0.5cm \protect\caption[] {(a) Bipartition of the Kitaev
     spin chain into subparts $A$ and $B$. (b) DMRG results
     for bipartite fluctuations as a function of the subsystem length
     $l_A = L$ at the critical point:
     $\mathcal{F}_{AB} = \alpha \ln L + \mathcal{O}(1)$ with
     $\alpha = 0.95/\pi^2$. The entanglement entropy is shown in the
     inset: $\mathcal{S} = (c/6) \ln L+ \mathcal{O}(1)$ with central
     charge $c=0.49$. (c) Bipartite fluctuations as a
     function of $\Delta = |J_2 - J_1|$ (here, $(J_1, J_2)<0$ such
     that the strong bonds occur on the $x$-link). The inset shows the
     correlation length $\xi = 0.20 \times \Delta^{-\nu}$ with
     $\nu = 0.94$. For the gapped phase, we set
     $L_{\text{total}}=1000$.}
    \label{fig:chain}
    \vskip -0.5cm
\end{figure}

%-------------------------------------------------------------------------------------------
\subsection{Results on fluctuations}

\subsubsection{Logarithmic growth}
 Below, we focus on $\mathcal{F}_{AB}$,
which defines the fluctuations between two subsystems (\ref{eq:fab})
associated to
bond-bond correlations (\ref{eq:bo}). 
Here the valence bond fluctuations $\mathcal{F}_{AB}$ also appear as the effective non-vanishing lower bound for mutual
information (\ref{eq:lb}).

Fig.~\ref{fig:chain} (a) depicts the bipartition we choose for the spin chain
with subsystem lengths
  $l_A = l_B = {L_{\text{total}}}/2 = L$. A
direct lattice summation in Appendix~\ref{app:lk1} leads to
   \begin{gather}
    \mathcal{F}_{AB} = \begin{cases} ({1}/{\pi^2}) \ln l_A + ({1}/{\pi^2}) \left( \gamma + \ln 2 - {1}/{2}\right), J_1 = J_2;\\
      c_1 \ln \xi + c_2 e^{-1} \left( 2 \xi^2 - \xi \right) + \mathcal{O}\left( 1    \right), \quad \ {|J_1|> |J_2|}.
      \label{eq:fc}
      \end{cases}
  \end{gather}
  with $\gamma \simeq 0.57721$
  the Euler constant. On the contrary, $\mathcal{F}_A$ always contains
  a higher order scaling linear in $l_A$.  
  
  Our findings (\ref{eq:fc}) are confirmed by
  DMRG simulations. At the gapless point $J_1 = J_2$ shown in
  Fig.~\ref{fig:chain} (b), we observe in $\mathcal{F}_{AB}$
   a logarithmic scaling with respect to the length of subregion $A$:
  $\mathcal{F}_{AB} \propto \ln l_A$. Roughly, one can identify this term by taking
  the 1D integral of the bond correlation $I(i,j)$ in Eq.~(\ref{eq:ik1}).
  Moreover, the pre-factor
  $\alpha/\pi^2$ is recovered with $\alpha = 0.95$. Here,
  the fact that $\alpha$ reproduces central charge $c = 1$ of the dual lattice,
  is in agreement with the free bond
  fermion representation~\cite{FS}.
  
  Fig.~\ref{fig:chain} (c) further probes the
  gapped region. When {$|J_1| \gg |J_2|$}, we check that
  $\mathcal{F}_{AB}$ goes to zero reflecting the crystallization of
  the dimers. Slowly closing the gap $\Delta$, 
  near the phase transition point, $|c_1| \gg |c_2|$. we check that 
 the logarithmic behavior $\propto \ln \xi$ dominates in $\mathcal{F}_{AB}$.

\subsubsection{Relation to entanglement entropy}
 
 It is interesting to go beyond the lower bound and reveal the relation between
 $\mathcal{F}_{AB}$ and the original entanglement entropy. Deep in the gapped phase
 driven by $|J_1| \gg |J_2|$, eigenstates are formed on strong $x$-links.
  $\mathcal{S}_A$ vanishes accordingly when the boundary is set on the weak
   $y$-link (see Fig.~\ref{fig:chain}, a). By increasing $|J_2|$, long-range entanglement emerges among the dimers, which is accompanied by a logarithmic growth in entropy associated to the correlation length
   \begin{gather}
     \mathcal{S}_A \propto \ln \xi. 
   \end{gather}
  The same response is observed in valence bond fluctuations $\mathcal{F}_{AB}$ of the gapped region.
  
 Meanwhile, the entropy reaches its maximum when the gap closes at 
 $J_1 = J_2$. Suppose the critical chain is finite with
 open boundaries, the entropy is proven to show the universal
 behaviour~\cite{cardy2004, affleck1991}:
   \begin{gather}
      \mathcal{S}_A = \frac{c}{6} \ln l_A + 2g + s_1, \label{eq:sk1}
   \end{gather}
 where $g$ counts the boundary
 entropy and $s_1$ stands for a non-universal constant. 
In inset of
 Fig.~\ref{fig:chain} (b), from DMRG, we check that the
 central charge extracted from the entropy (\ref{eq:sk1}) reads: $c = 0.49\sim 1/2$.
 It can be understood from the fact that after the Jordan-Wigner transformation, half of
 the spin degrees of freedom are disentangled from the Hamiltonian (\ref{eq:hc0}) by decoupling all
 $d$-Majorana fermions.

 At the gapless point, both $\mathcal{F}_{AB}$ and $\mathcal{S}_A$ then share a logarithmic
 growth with subsystem size typical of (critical) conformal field
 theories in one dimension. Related to this finding, we would like to address the following comment: to evaluate the valence bond fluctuations
 we diagonalize the spectrum in momentum space in the $\psi$ basis,
 whereas the entanglement entropy reflects the real space degrees of
 freedom on the original lattice. This justifies why in our
   calculations the central charge $c=1$ is revealed in the valence
 bond fluctuations, whereas the central charge $c=1/2$ is observed in
 the entanglement entropy.
 
 We establish that in both the gapped and gapless phases of one-dimensional Kitaev spin liquids, 
 there is an identical scaling rule between the valence bond
 fluctuations and the entanglement entropy
   \begin{gather}
     \mathcal{F}_{AB} \sim \mathcal{S}_A.
   \end{gather}

%-----------------------------------------------------------------------------------------------------------------------------------------------------
 \section{Model on the honeycomb lattice} 
 \label{sec:honeycomb}
    As discussed briefly in Sec.~\ref{sec:lowerbound} and as will be described below,
 on the two-dimensional
 honeycomb lattice, due to the protection of $0$-flux configurations
 in the ground state, the two-spin correlator also vanishes beyond
 nearest neighbors in all phases~\cite{baskaran2007}. The valence bond
 correlator, however, preserves flux pairs in neighboring plaquettes
 and supports gapless fermion excitations. Although Ref.~\cite{shuo2008} finds
 numerically that it exhibits a power-law decay in the gapless phase and an exponential decay in gapped
 phases,  the valence bond correlator itself does not locate
   precisely the phase transition (see Fig.~\ref{fig:bb_r}, a). It
   motivates us to develop an approach of
    evaluating its global fluctuations on a bipartite
   lattice. The enhanced features in fluctuations come intrinsically from the spatial dependence and 
   anisotropy of the local bond correlator. 
   
   We demonstrate below that the valence bond
   fluctuations and the entanglement entropy allow us to locate 
 quite accurately the phases and quantum phase transitions in the 
   two-dimensional Kitaev honeycomb model~\cite{kitaev2006}. 
   
    \begin{figure}[t]
   \begin{center}
                \includegraphics[width=.45\linewidth]{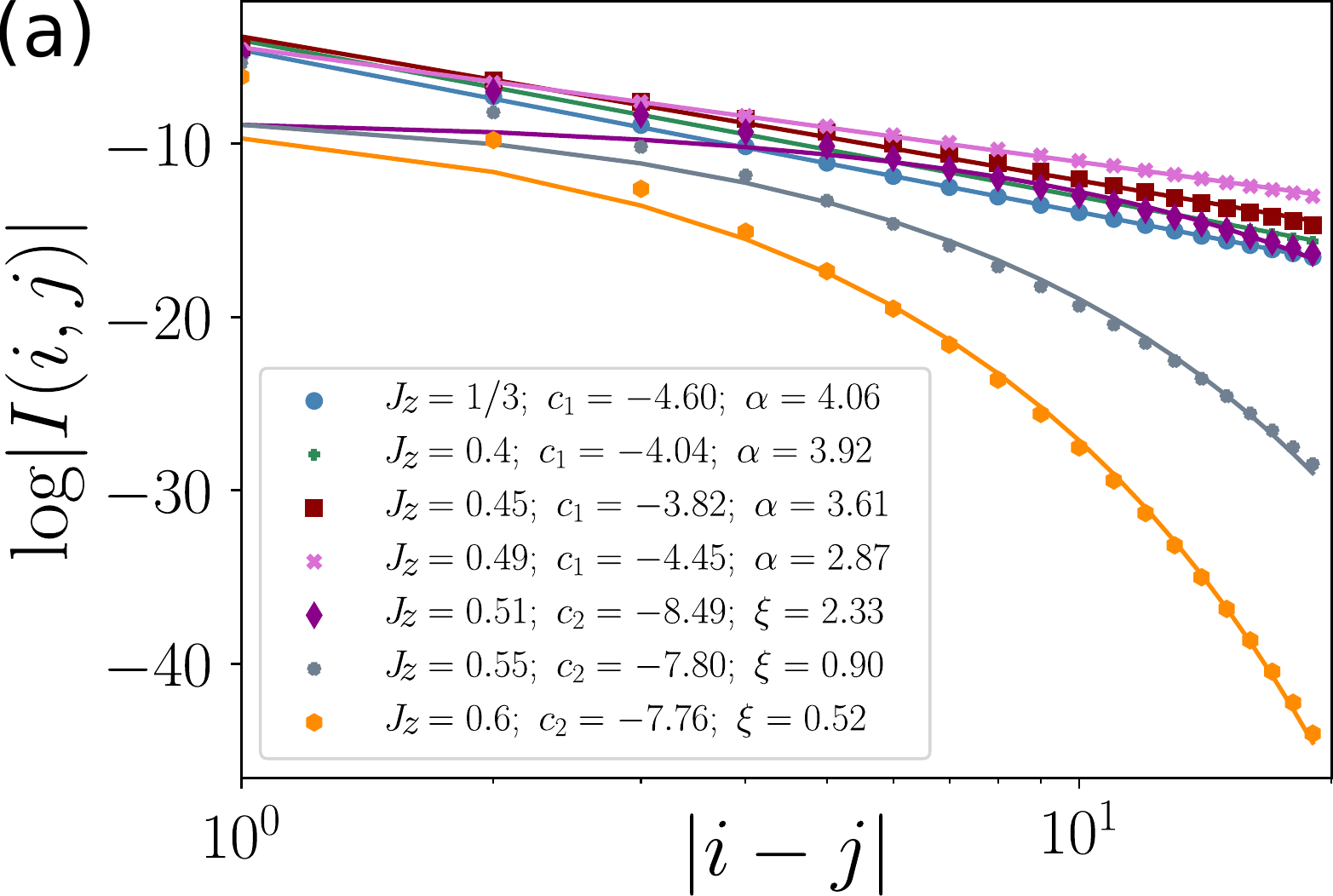}   \ 
                   \includegraphics[width=.45\linewidth]{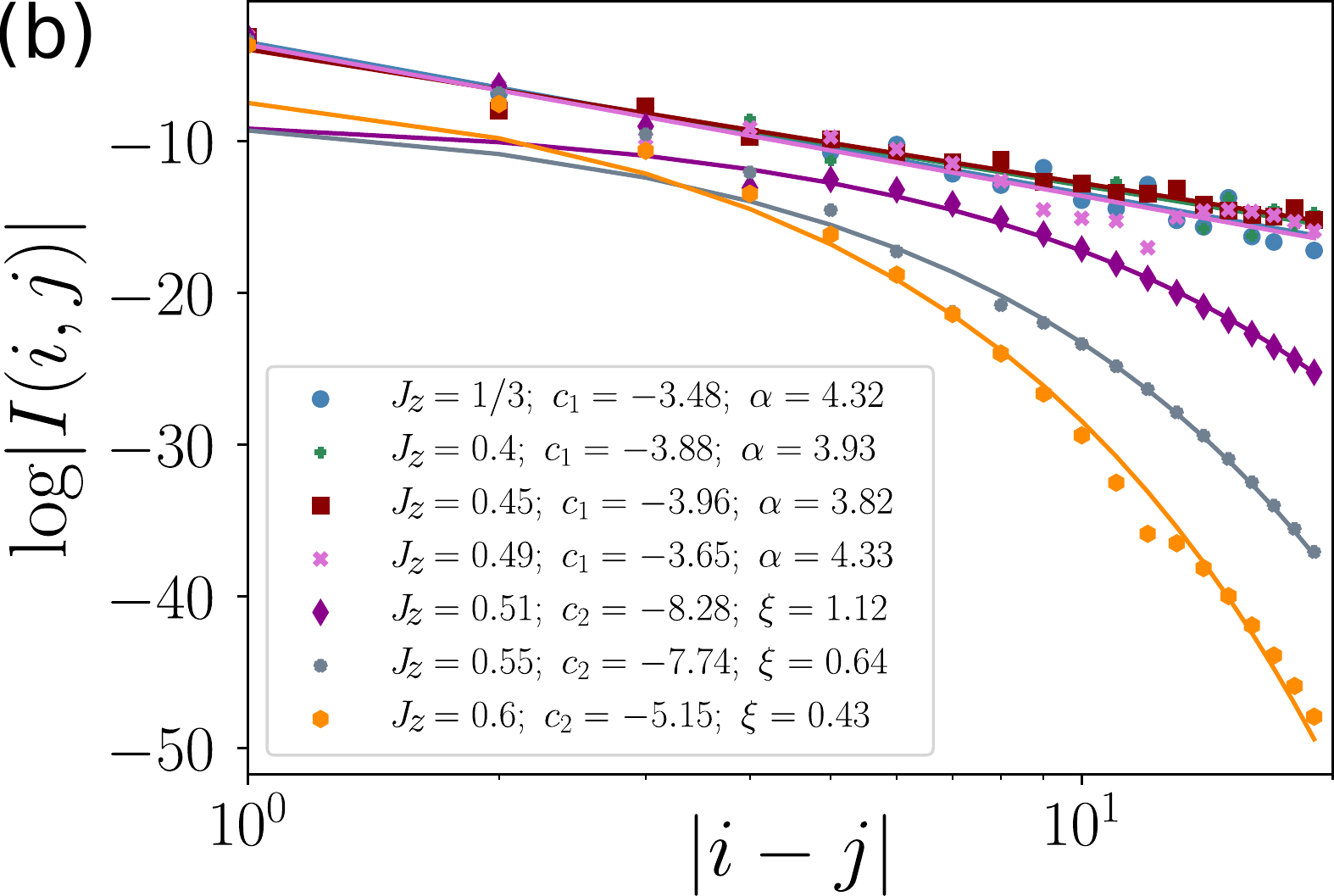} \\ \vspace{0.2cm}
                   \includegraphics[width=.45\linewidth]{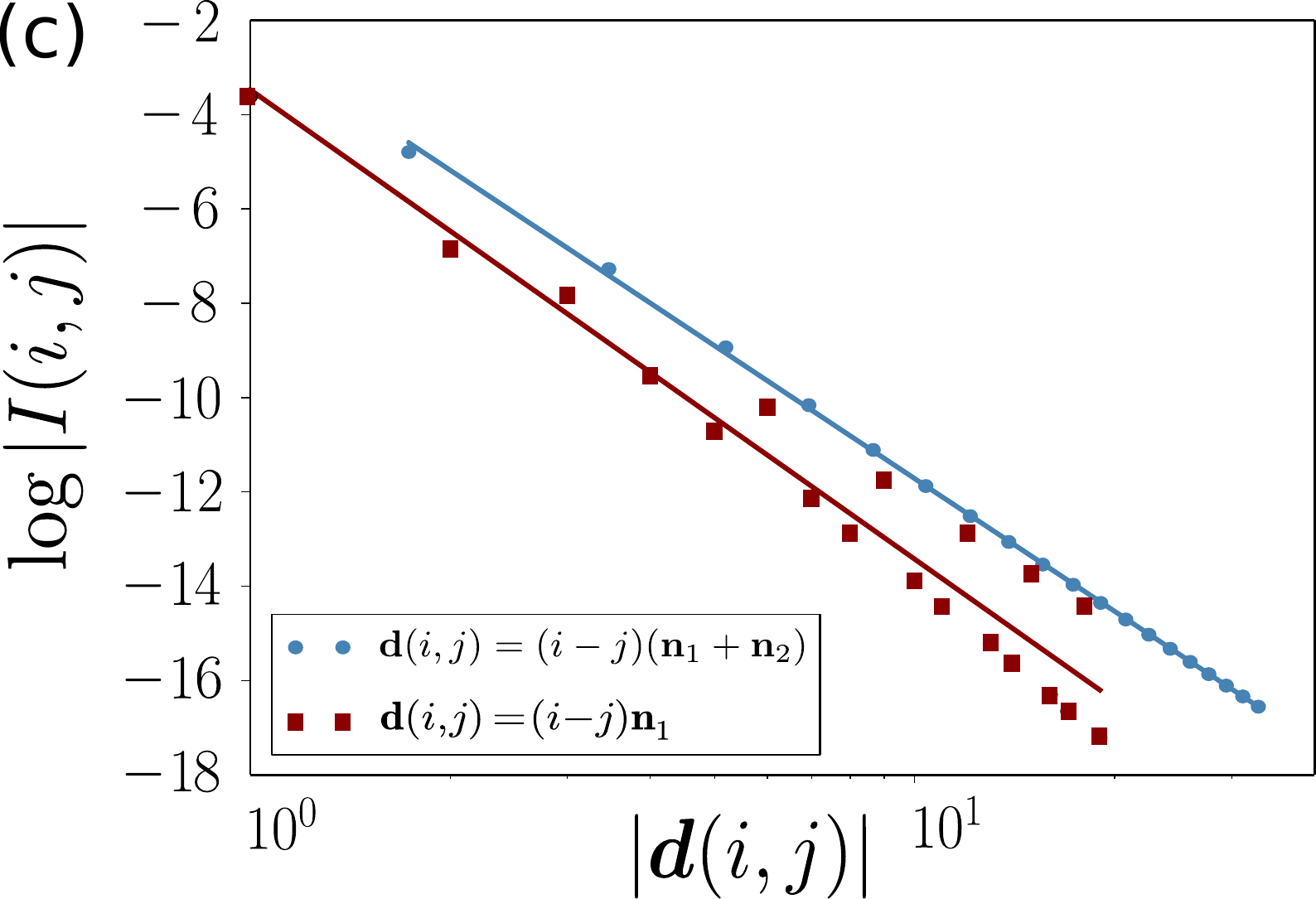} \ 
                  \includegraphics[width=.45\linewidth]{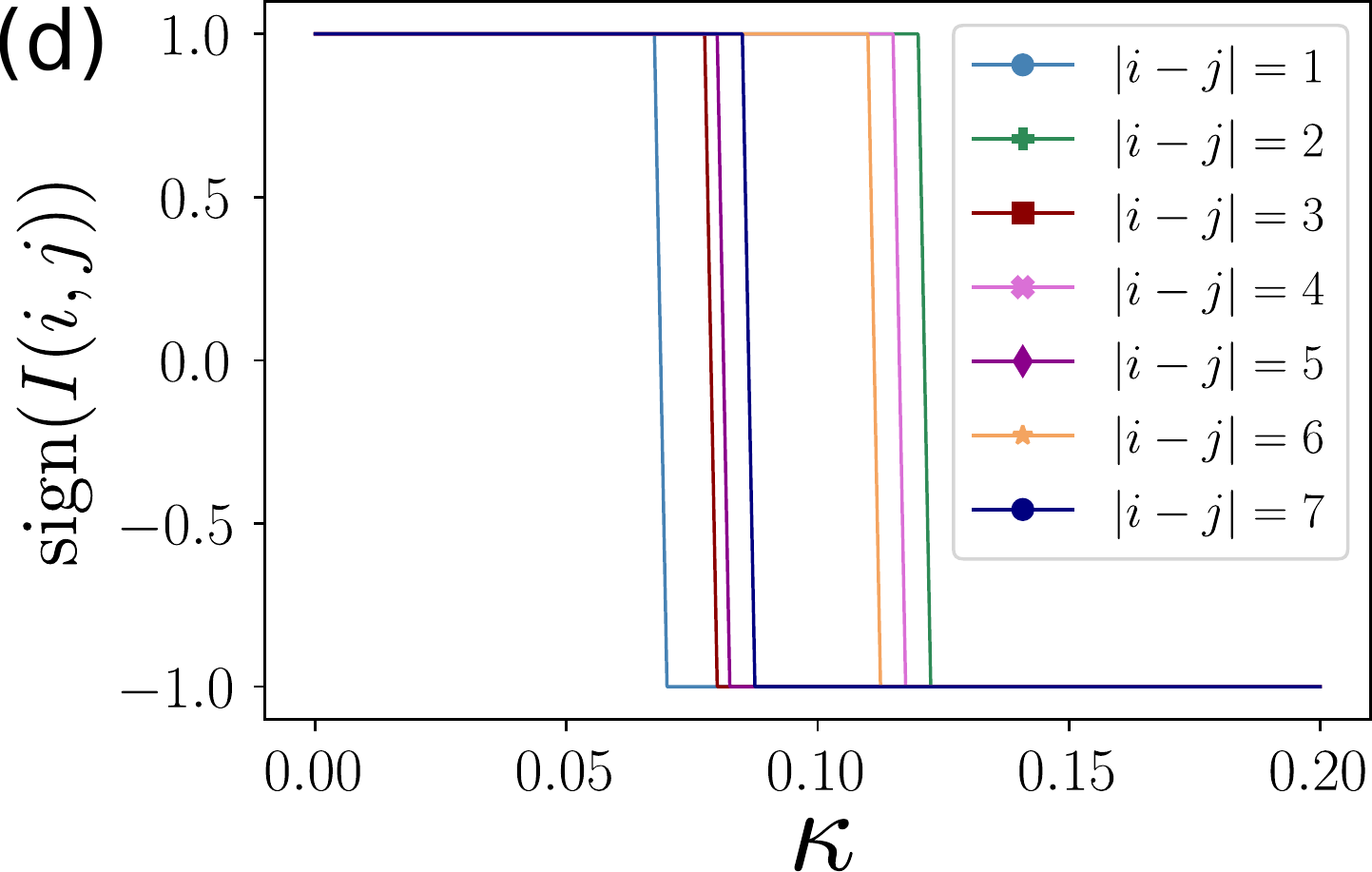} \\
                   \vspace{0.2cm}
                 \includegraphics[width=.45\linewidth]{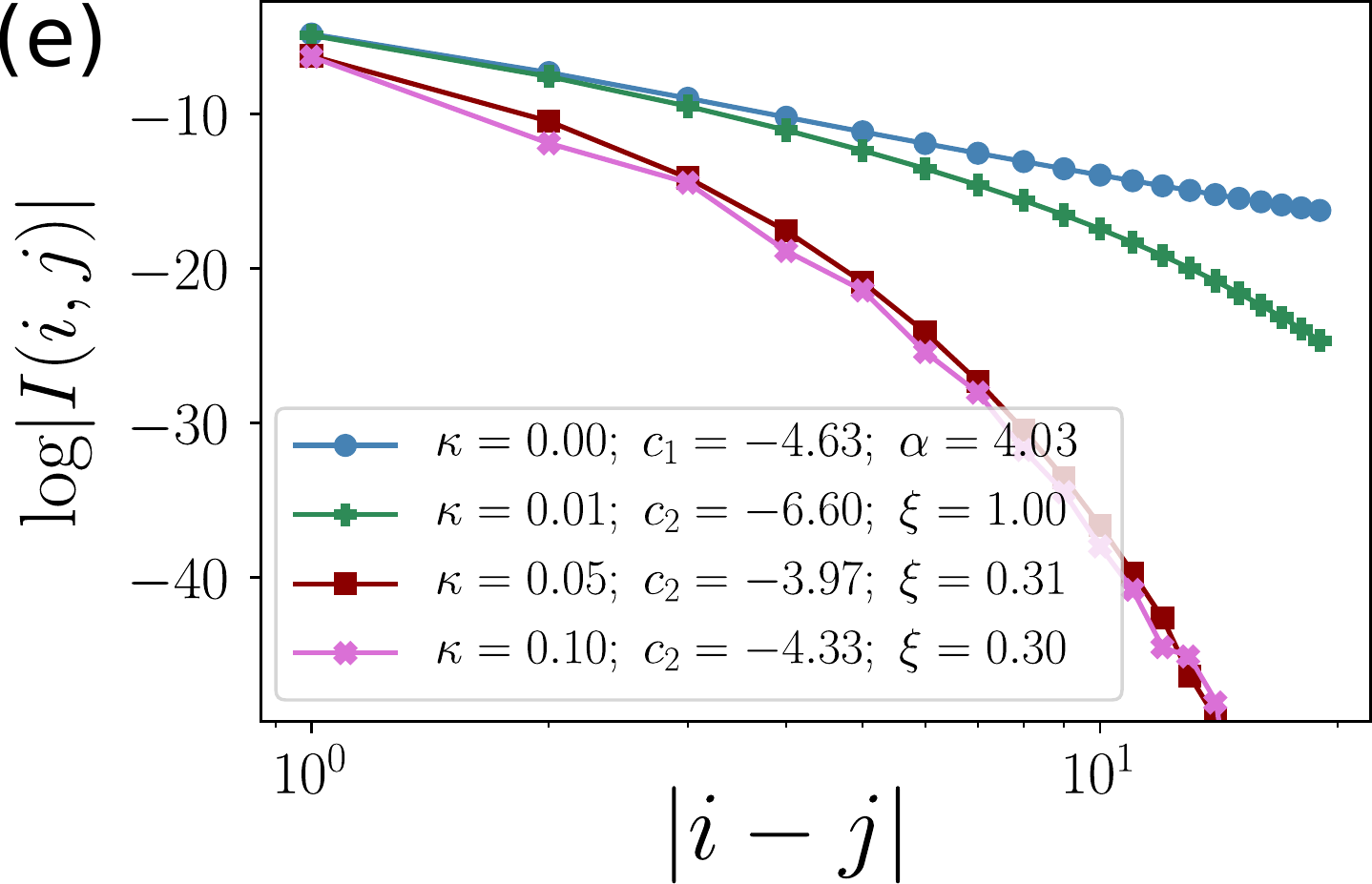} \ 
                 \includegraphics[width=.45\linewidth]{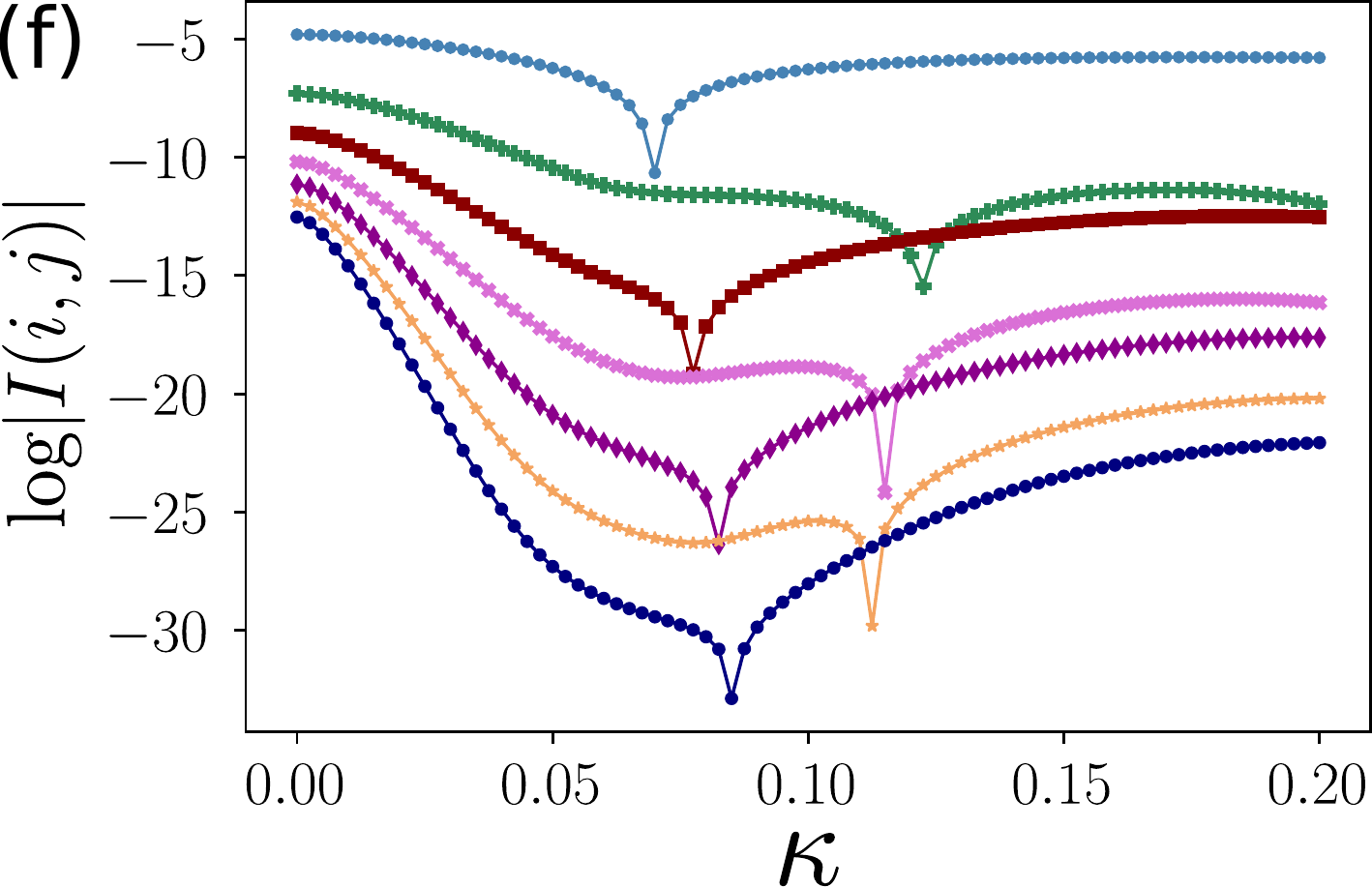}  
   \end{center}
    \vskip -0.5cm \protect\caption[]
    {(a), (b) and (c): valence bond correlation functions for different phases in the pure Kitaev honeycomb model. The coupling constants are chosen according to $J_x = J_y, \sum_a J_a = 1$ and we take the total system size as $N = L \times L = 100 \times 100$. While the gapped phase ($J_z > 0.5$) exhibits an exponential decay in bond correlators: $\log |I (i,j)| = c_2 -|i-j|/\xi$, the gapless intermediate phase ($J_z \le 0.5$) supports a power-law decay: $\log |I (i,j)| = c_1- \alpha \log |i-j|$. The relative vector is set along the direction $\textbf{d}(i,j) = \vec{r}_i - \vec{r}_j$: (a) $\textbf{d}(i,j) = (i-j)(\vec{n}_1 + \vec{n}_2) = (0,\sqrt{3}(i-j))$; (b) $\textbf{d}(i,j) = (i-j)\vec{n}_1 = ((i-j)/2,\sqrt{3}(i-j)/2)$. (c) Anisotropy effects in the gapless phase ($J_x = J_y = J_z = 1/3$). Valence bond correlator for the intermediate phase in the presence of a uniform magnetic field $[111]$: (d) Sign change; (e) Exponential decay for the finite field; (f) Absolute amplitude with varied magnetic strengths. Here we present the case $J_x = J_y = J_z = 1/3$. The direction of the relative vector is chosen on  $\textbf{d}(i,j)  = (i-j)(\vec{n}_1 + \vec{n}_2)$.}
	\label{fig:bb_r}
	\vskip -0.5cm
\end{figure}

 %----------------------------------------------------------------------------------------------------------------------------------------------------
\subsection{Majorana representation}

We start from the Hamiltonian $\mathcal{H} = -\sum_{\langle ij \rangle_a} J_a \sigma_i^a \sigma_j^a$,
under the perturbation $\mathcal{V} = -\sum_j h_a \sigma_j^a$, where
$\langle ij \rangle$ represents two nearest-neighbor sites, forming the bonds and $a = x, y, z$
denotes one of three different Ising couplings assigned onto them
(see Fig.~\ref{fig:2d0}, a).  When $|h_a| \ll |J_a|$, the
{cubic} term in perturbation theory breaks time-reversal symmetry, and
the effective Hamiltonian is simplified to~\cite{kitaev2006}
  \begin{gather}
    \mathcal{H} = -\sum_{\langle i j \rangle_a} J_a \sigma_i^a \sigma_j^a - \kappa \sum_{\langle  \langle ik \rangle \rangle} \sigma_i^a \sigma_j^c \sigma_k^b.
    \label{eq:hh}
  \end{gather}
  Here $\kappa \sim h_xh_yh_z/(J_aJ_b)$ and $\langle \langle ik \rangle \rangle$ describes next-nearest neighbors $i$ and $k$ connected by site $j$. Below, we study the effect of $\kappa$ in the perturbative regime of Eq. (\ref{eq:hh}) where $\kappa\ll J_x+J_y+J_z$.
  
  On the honeycomb lattice, one can write spin operators in terms of four Majorana fermions per site~\cite{kitaev2006}: 
   \begin{gather}
     \sigma_j^a = ic_jc_j^a,
   \end{gather}
   with the constraint for a physical state    
   \begin{gather}
     D_j = c_j c_j^x c_j^y c_j^z = 1.
    \end{gather}
   Here, $c_j$ are matter Majorana fermion operators determining the fermionic spectrum and $c_j^a$ belong to
   gauge Majorana fermions which in the ground state, can be fixed artificially. Noticing that $[ic_i^ac_j^a, \mathcal{H}] = 0$
   and $(ic_i^ac_j^a)^2= 1$,
we choose the gauge $u_{\langle ij \rangle_a} = ic_i^ac_j^a := + 1$ for $i$ in sublattice
  $\{ 1 \}$, as depicted in {Fig.~\ref{fig:2d0} (a)}, such that the total flux in the plaquette becomes zero
  consistent with Lieb's theorem~\cite{lieb1994}. The definition of two sub-lattices, named $1$ and $2$ in {Fig.~\ref{fig:2d0} (a)}, follow the one-dimensional situation and the presence of two inequivalent bonds
  in a (zig-zag) given direction. 
  
   \begin{figure}[t]
    \begin{center}
      \includegraphics[width=0.22\textwidth]{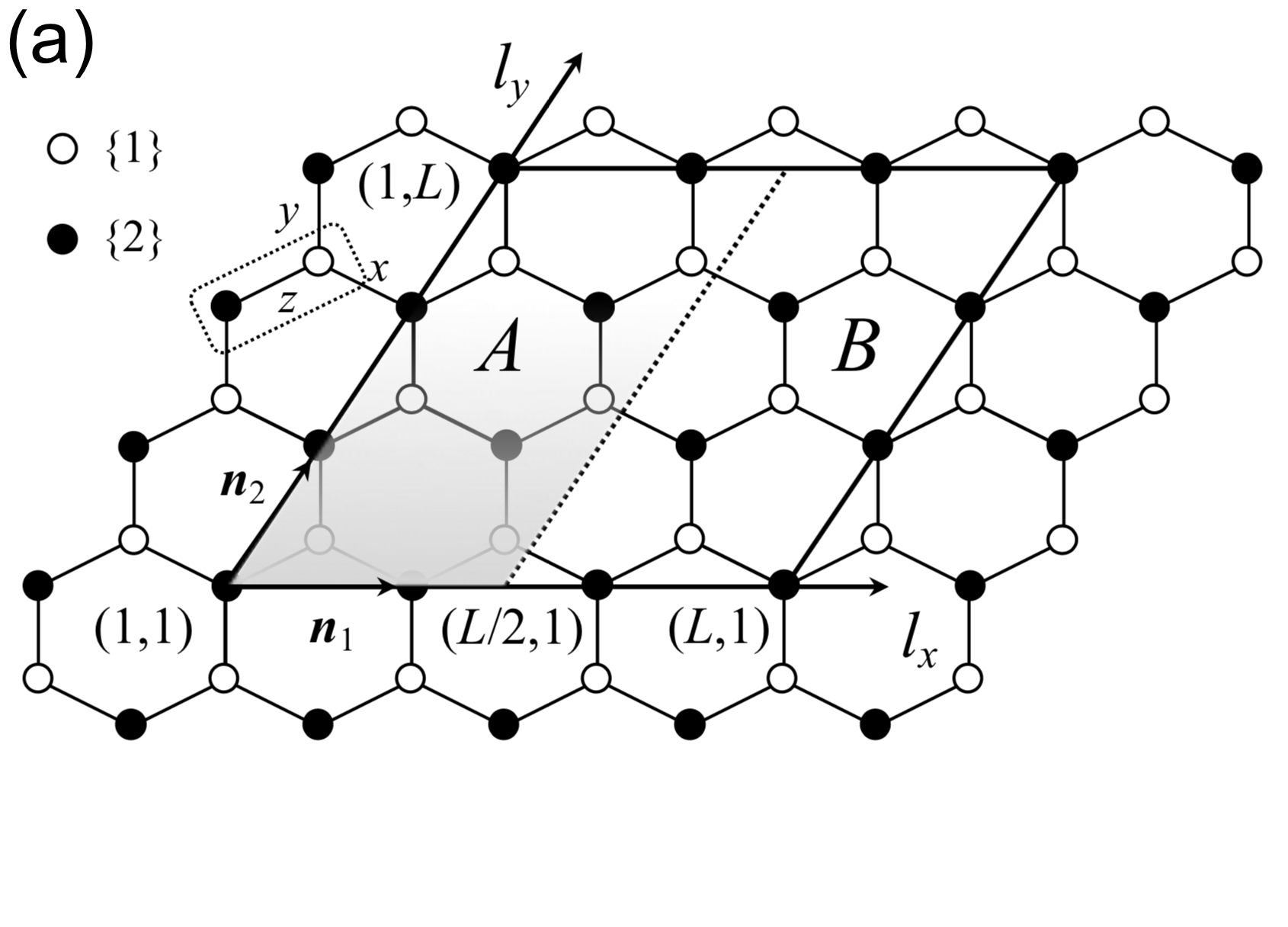} \includegraphics[width=0.225\textwidth]{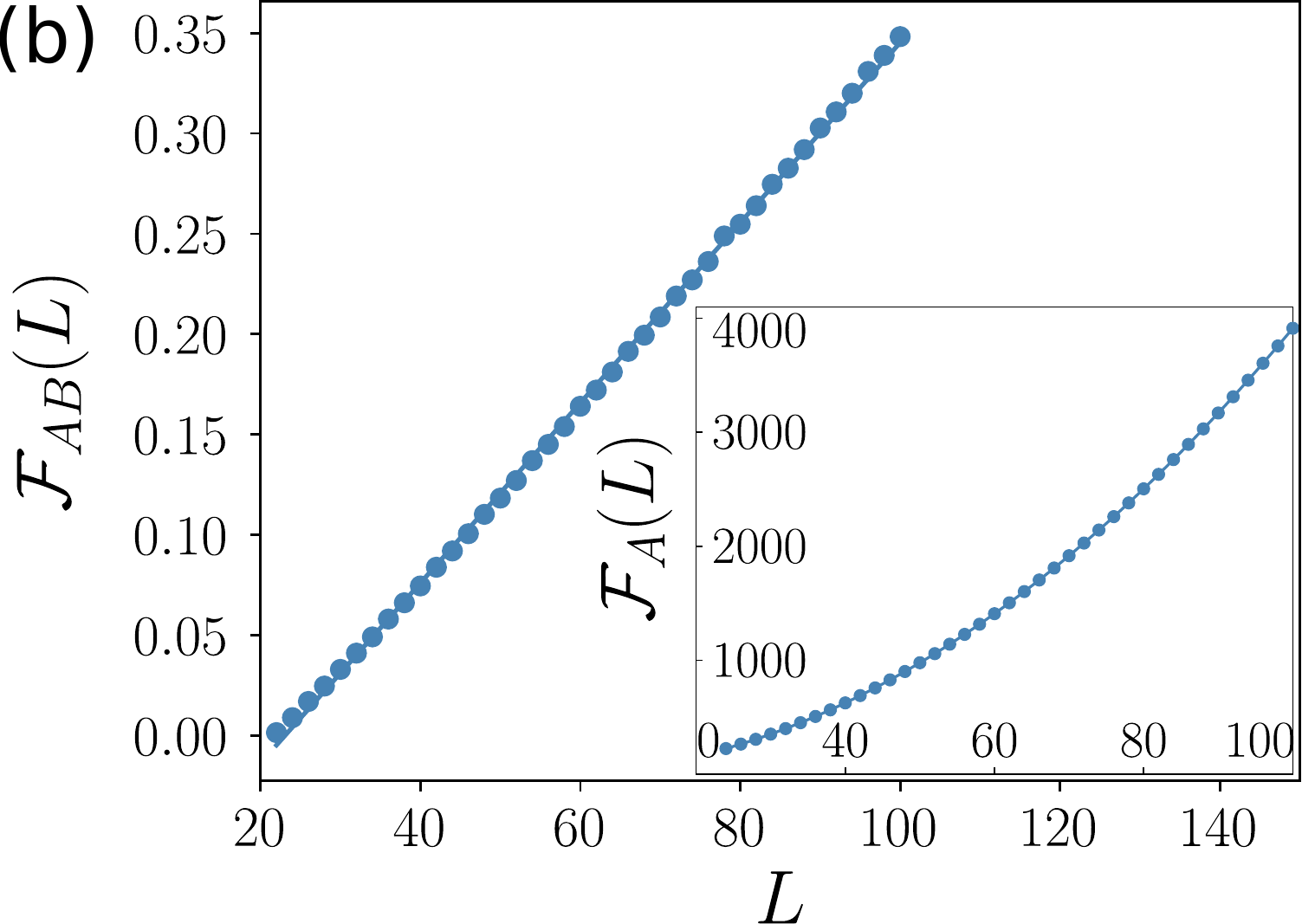}  
    \end{center}
  \vskip -0.6cm \protect\caption[] {(a) Bipartition into two
    subsystems $A$ and $B$. The parallelogram is formed with unit
    vectors: $\vec{n}_1 = (1/2, \sqrt{3}/2)$,
    $\vec{n}_2 = (-1/2, \sqrt{3}/2)$. The bond observable is chosen on
    each $z$-link (dashed box); (b) Scaling of
    $\mathcal{F}_{AB} = {\alpha}_{\mathcal{F}} L + \mathcal{O}(\ln L)$
    in the gapless intermediate phase $J_x = J_y = J_z$. We obtain
    numerically ${\alpha}_\mathcal{F} = 0.00449$. The inset shows the
    scaling of
    $\mathcal{F}_A = \alpha L^2 + \beta L + \mathcal{O}(\ln L)$ with
    $\alpha = 0.391, \beta = 0.0129$.}
    \label{fig:2d0}
    \vskip -0.5cm
\end{figure}

 The effective Hamiltonian~\cite{kitaev2006} then presents a quadratic form in terms of $c_j$ as in
  one dimension (\ref{eq:hc0}):
\begin{equation}
\mathcal{H} = \sum_{\langle ij \rangle_a} i J_a c_i c_j + \sum_{\langle \langle ik\rangle \rangle} i\kappa c_ic_k.  
\end{equation}
To obtain the eigenvectors, again it is more convenient to use the
basis of complex bond fermions:
 \begin{gather}
   \psi_{r} = \frac{1}{2}(c_{r, 1} + ic_{r, 2}) \label{eq:hbfer}
 \end{gather}
with $r$ the unit cell
coordinate defined on the $z$-links and $\{1,2\}$ the sublattice
index.  
In momentum space, one arrives at a similar form of Hamiltonian as
Eq. (\ref{eq:hkc}) with matrix elements
  \begin{gather}
    \xi_{\vec{k}} = \mathfrak{Re}{f(\vec{k})}, \quad
    \Delta_{\vec{k}} = -g(\vec{k})+ i \mathfrak{Im} f(\vec{k}). \label{eq:xid}
  \end{gather} 
Here,
$f(\vec{k})$ and $g(\vec{k})$ are functions of the form
  \begin{gather}
  f(\vec{k}) = J_x e^{i\vec{k}\cdot \vec{n}_1} + J_y e^{i\vec{k}\cdot
  \vec{n}_2} + J_z, \notag \\
  g(\vec{k}) = 2\kappa [ \sin(\vec{k}\cdot \vec{n}_1) +
\sin(\vec{k}\cdot ( \vec{n}_2 - \vec{n}_1) - \sin(\vec{k}\cdot
\vec{n}_2) ] \label{eq:fg0}
  \end{gather}
with two unit vectors
$\vec{n}_1 = (1/2, \sqrt{3}/2)$ and 
    $\vec{n}_2 = (-1/2, \sqrt{3}/2)$ shown in Fig.~\ref{fig:2d0} (a).
    We check that the energy spectrum obtained from Eqs.~(\ref{eq:ev}), (\ref{eq:xid}) and (\ref{eq:fg0})
reveals a gapless intermediate semi-metal phase ~\cite{kitaev2006}
   \begin{gather}
      |J_a| \le |J_b| + |J_c|
   \end{gather}
where $(a,b,c)$ involve all permutations of
$(x, y, z)$.

%----------------------------------------------------------------------------------

\subsection{Valence bond correlator}
\label{sec:vbc}
For the honeycomb lattice, we choose to measure the valence bond correlator 
onto the $z$-links, as compared to the $x$-links used for the one-dimensional chain in Sec. \ref{sec:chain}:
 \begin{gather}
    Q_i = \sigma_{i,1}^z \sigma_{i,2}^z = -ic_{i,1}c_{i,2} u_{\langle i_1 i_2 \rangle_z} := -ic_{i,1}c_{i,2}. \label{eq:bc}
 \end{gather}
Since any physical observable is independent of a specific gauge~\cite{baskaran2007}, in the last equality, we can adapt our gauge choice for the gauge Majorana fermions. 
One thus sees the bond correlator probes the matter Majorana fermions without disturbing the gauge part.

On the contrary, a single spin operator $\sigma_{j,1}^z = ic_{j,1}c_{j,1}^z$ influences the gauge structure~\cite{baskaran2007}.
Defining the bond gauge fermion $\chi = (c_{j,1}^z + i c_{j,2}^z)/2$, we find that the number operator
   $ N_\chi = \chi^\dagger \chi = (u_{\langle j_1 j_2 \rangle_z} + 1)/2$ takes the value $1$ and $0$ depending on the gauge choice of $u_{\langle j_1 j_2 \rangle_z} = \pm 1$.
 In either gauge,     $\sigma_{j,1}^z = ic_{j,1}(\chi^\dagger + \chi)$ changes the occupation number of the bond gauge fermion $\chi$.
 It is equivalent to flip the linking number $u_{\langle j_1 j_2 \rangle_z}$ to $-u_{\langle j_1 j_2 \rangle_z}$ and excite one $\pi$-flux pair in two neighbouring plaquettes.
 Therefore, the two-spin correlation is totally suppressed by the static $\mathbb{Z}_2$ gauge background beyond nearest neighbours (the latter case goes back to the measurement on a single bond).

Next, we focus on the local structure of the non-vanishing valence bond correlation.  From
  Wick's theorem, one obtains a compact form
  \begin{gather}
      I(i,j) = -\sum_{\vec{k},\vec{q}}
    \frac{e^{i(\vec{k}+\vec{q})\cdot(\vec{r}_j-\vec{r}_i)}[g(\vec{k})g(\vec{q})
      + f(\vec{k})^*f(\vec{q})]}{N^2
      \sqrt{(\xi_{\vec{k}}^2+|\Delta_{\vec{k}}|^2)(\xi_{\vec{q}}^2+|\Delta_{\vec{q}}|^2)}},
    \label{eq:bb_k}
  \end{gather}
with $N$ the total number of lattice unit cells.

 %---------------------------------------------------------------------------------------------------------------
 \subsubsection{Zero field}
 
In the absence of magnetic field $\kappa = 0$, $I(i, j)$ has no singularities in the three gapped Abelian phases, therefore this results in an exponential decay of $I(i,j)$. In the intermediate gapless {semi-metal} phase, singularities appear at two Dirac points $\pm \vec{k}^*$.
   
 We first look at the behavior of bond correlations in the gapless region. A detailed analysis of the asymptotic behavior of $I(i, j)$ at long distances can be found in Appendix~\ref{app:kitaev}.
 Performing an expansion around the two Dirac points similar to the $p_x+ip_y$ superconductor~\cite{LoicPRB}, we recover a power-law decay~\cite{shuo2008} 
  \begin{gather}
     I(i,j) = \frac{\widetilde{c}_1}{r^{4}}, \label{eq:i_func}
   \end{gather}  
     and establish that the $\widetilde{c}_1$ coefficient depends on the cutoff function $t(\Lambda)$  and on the anisotropic function $Y(\vec{r})$. More precisely,
    \begin{gather}
         \widetilde{c}_1 = t^2(\Lambda) \cdot Y(\vec{r}), \notag \\
          Y(\vec{r}) = \cos^2(\vec{k}^*\cdot \vec{r}) - \cos^2(\theta^*).   \label{eq:yfunc}
    \end{gather}
    Here, $\theta^*$ is the angle between the vectors
    $\vec{r}=\vec{r}_i-\vec{r}_j$ and $\vec{k}^*$. The space variable
    $r$ refers to $| \vec{r}_i-\vec{r}_j |$. 
    
We can start from the simplest case by making the directions of  $\vec{r}$ and $\vec{k}^*$ perpendicular to each other: $\vec{r}_\perp =\vec{r}_j - \vec{r}_i = (j-i)(\vec{n}_1 + \vec{n}_2)$. The spatial oscillations disappear in the bond correlator with $Y(\vec{r}_\perp) = 1$.  Our analytic expression becomes consistent with the numerical fitting results of Ref.~\cite{shuo2008}. 
Shown in Fig.~\ref{fig:bb_r} (a), it supports a smooth curve of $I(i,j)$ revealing  the $r^{-4}$ scaling in the gapless region ($J_z < 0.5$).

We can derive a more precise analysis. At the gapless point $J_x = J_y  = J_z =1/3$, in Appendix.~\ref{app:kitaev} we derive the expression for  the cutoff function
     \begin{gather}
       t(\Lambda) = \frac{\sqrt{3} }{2\pi}   \int_{0}^{\Lambda} J_1(k)kdk, \label{eq:cfunc}
    \end{gather}
where inside the integral $J_1(k)$ denotes the Bessel function of the first kind. Here, the cutoff $\Lambda = \xi r$ can be further approximated by setting the radius of the momentum integration $\xi = 1$ and taking $r \simeq r_{\text{max}}$ to be the total system size $L = 100$. Analytically, we obtain $\log |I (i,j)| = c_1- \alpha \log |i-j|$ with $c_1 = \left. \log (\widetilde{c}_1/9)\right|_{\Lambda = 100} = -4.63$, $\alpha = 4$. Here, $``\log"$ is equivalent to the natural logarithm with the base ${e}$.  It recovers well the numerical fitting result (see Fig.~\ref{fig:bb_r}, a):  $c_1 = -4.60, \alpha = 4.06$. 

It is important to stress that Ref.~\cite{shuo2008} has not  pointed out the role of the anisotropic $Y$-function. Once shifted to other directions $Y(\vec{r}) \ne $ cst, $c_1 =  \log ( |\cos^2(\vec{k}^*\cdot \vec{r}) - \cos^2(\theta^*)|) - 4.63.$ Accordingly, as verified by Fig.~\ref{fig:bb_r} (b) and (c), the sampling points of  $\log |I(i,j)|$ along the non-perpenticular direction oscillate rapidly.  
{We also emphasize here  the forms of $\tilde{c}_1$ and the anisotropic $Y$-function in  Eq.~(\ref{eq:yfunc}) remain true for the whole gapless region.} {Later, we will study these anisotropic effects on the bipartite fluctuations in relation with Fig.~\ref{fig:2d}.

For the gapped phase, on the other hand, {from numerics} $I(i,j)$ follows an exponential decay with a fast decreasing correlation length shown in Fig.~\ref{fig:bb_r} (a). Meanwhile, in Fig.~\ref{fig:bb_r} (b), one observes less anisotropy effects in the gapped region.  

It may be relevant to mention that once the gapless intermediate phase is subject to a magnetic field along the $\hat{z}$ direction, an identical power-law behavior (including
 the same angular dependence) emerges in the dynamical correlation
 function~\cite{kitaev2011}:
  \begin{align}
    &g(t,\vec{r}) = \langle \sigma_r^z(t) \sigma_0^z(0)\rangle - \langle \sigma_0^z(0)\rangle^2 \notag \\
    &\simeq \frac{64h_z^2}{\pi^2 h_0^2}  \frac{r^2(\cos^2(\vec{k}^*\cdot \vec{r}) - \cos^2(\theta^*)) -3(Jt)^2 \cos^2(\vec{k}^*\cdot \vec{r})}{(r^2-3(Jt)^2)^3}
  \end{align}
  where $h_z$ is the strength of the magnetic field and the parameter $h_0$ can be estimated as $h_0 \sim J$. We find at long distances and at a
 finite time,
   \begin{gather}
    t \ll r \to \infty, \quad  \frac{g(t,\vec{r})}{I(\vec{r}, \kappa = 0)} \propto \text{cst}.  
  \end{gather}
    Both observables are proportional to the density-density correlation
 function of the bond fermions $\psi_r$ in (\ref{eq:hbfer}).
 
 %-------------------------------------------------------------------------------------
 \subsubsection{Small finite field}

We further study the effects of a small uniform magnetic field on the bond correlation in
the intermediate phase.
For simplicity, we take $J_x = J_y = J_z = J = 1/3$. 

When {$0 < \kappa \ll J $},
 a gap opens and the valence bond correlator
 in Fig.~\ref{fig:bb_r} (e) now reveals an exponential decay, 
 similar to three gapped spin liquid phases. 
 Yet its sign changes from positive to negative when
 increasing the strength of the magnetic field
  (see Fig.~\ref{fig:bb_r}, d). Consequently, in Fig.~\ref{fig:bb_r} (f) we observe an enhancement in the amplitude of bond correlation functions once the magnetic field is sufficiently large. 
  
  We find that this sign change originates from the competition between the Ising interactions and the external magnetic field. 
For $\kappa \ne 0$, the valence bond correlator (\ref{eq:bb_k}) can be expressed in an alternative form
  \begin{align}
    I(i,j) &= F(\vec{r})^2 - G(\vec{r})^2, \notag \\
    F(\vec{r}) &= \frac{1}{N} \sum_k e^{i\vec{k}\cdot{\vec{r}}} \frac{f(k)^*}{\sqrt{\xi_k^2 + |\Delta_k|^2}}, \notag \\
    G(\vec{r}) &= \frac{1}{N} \sum_k 
                       e^{i\vec{k}\cdot{\vec{r}}} \frac{g(k)}{\sqrt{\xi_k^2 + |\Delta_k|^2}}.
    \label{eq:i_fg}
  \end{align}
 While the Ising interactions give a positive contribution to the bond correlators, the external magnetic field gives a negative one. 
 
Changing the strength of the external magnetic field $\kappa$, it is verified that  
  \begin{gather}
    \quad \text{Sign}\left( \frac{\partial I(i,j)}{\partial \kappa} \right) = \text{Sign}(-\kappa),
  \end{gather}
  as $ {\partial_\kappa I(i,j)} = 2 [ {\partial_\kappa F(\vec{r})} \cdot F(\vec{r}) - {\partial_\kappa G(\vec{r})} \cdot G(\vec{r})]$ and
  \begin{align}
     \frac{\partial F(\vec{r})}{\partial \kappa} &= -\frac{1}{N} \sum_k e^{i\vec{k}\cdot{\vec{r}}} \frac{f(k)^*g(k)^2}{\kappa(\xi_k^2 + |\Delta_k|^2)^{3/2}}, \notag \\
       \frac{\partial G(\vec{r})}{\partial \kappa} &= \frac{1}{N} \sum_k e^{i\vec{k}\cdot{\vec{r}}} \frac{g(k) |f(k)|^2}{\kappa(\xi_k^2 + |\Delta_k|^2)^{3/2}}.
   \end{align}   
When $\kappa > 0$, the derivative of $I(i,j)$  is always negative. 

The monotonically decreasing bond correlation function is expected to cross zero around the point where the strengths of the Ising interactions and the magnetic field are comparable. 
We can roughly estimate the crossover point by starting from a relatively small $\kappa$ parameter.
In this circumstance,  $I(i,j)$ is still governed by an expansion $|\delta \vec{k}| \in \Omega(0, 1)$ around two  original Dirac points $\pm \vec{k}^*$. The denominator in Eq.~(\ref{eq:i_fg}) turns out to be
  \begin{gather}
    E_k = \sqrt{\xi_k^2 + |\Delta_k|^2} = 3\sqrt{3} |\kappa| \sqrt{1+(\lambda k)^2}.
  \end{gather}
  When the parameter $\lambda = J/(6\kappa) > 1$, $\kappa < \kappa_c = 0.055$, the $F(\vec{r})^2$ term arising from the Ising interactions is dominant and $I(i,j)$ keeps a positive sign. Otherwise $\lambda \ll 1$, $\kappa \gg \kappa_c$, then the $-G(\vec{r})^2$ term from the external magnetic field grows steadily and has the tendency to drive $I(i,j)$ negative. In accordance with the numerical calculations shown in Fig.~\ref{fig:bb_r} (d) and (f), for different distances, all crossover points where $I(i,j)$ changes sign
  are located at $\kappa > \kappa_c$.

%---------------------------------------------------------------------------------------
\subsection{Results on fluctuations}
  \label{sec:bfha}
    \subsubsection{Area law}
    
{Next, to gain some intuition on the behavior of bipartite fluctuations,
in Appendix~\ref{app:las}
 we
perform analytically the lattice summation by assuming an isotropic form
of {$I(i,j)$}, namely with $Y(\vec{r}) = 1$.

Given a bipartition on the honeycomb lattice 
represented in Fig.~\ref{fig:2d0} (a),
we first derive a general scaling form for fluctuations within an arbitrary region  $\Omega = l_x \times l_y$
   \begin{gather}
     I(r) \propto \frac{1}{r^\alpha}, \quad \mathcal{F}_{\Omega} \propto 
       \begin{cases}
         L^4, \quad \alpha = 0; \\
         L^3, \quad \alpha = 1; \\
         L^2, \quad \alpha \ge 2,
       \end{cases} \label{eq:gsl0}
   \end{gather}
with $l_x$ and  $l_y$ of the same order as $L$.

For the Kitaev honeycomb model, from Sec.~\ref{sec:vbc} we see the valence bond correlator reveals
a power law decay ($\alpha = 4$) in the gapless phase and an exponential
decay ($\alpha \to \infty$) in the gapped phases. Therefore,  in all phases
$\mathcal{F}_{A}$ shows the volume law: $\mathcal{F}_{A} \propto L^2 = \mathcal{V}$. 
As usual, we can extract $\mathcal{F}_{AB}$ from the equality (\ref{eq:ffab}):  $ \left| \mathcal{F}_{A \cup B} - \mathcal{F}_A - \mathcal{F}_B \right|/2$.
With a subsystem size
$A = B = {(L/2)} \times L$, it is noticeable that 
the volume term vanishes after the subtraction, leading to an area law in $\mathcal{F}_{AB} \propto L = \mathcal{A}$, where $\mathcal{A}$ refers then to an area.

Meanwhile, under the $Y$-isotropic form assumption, we establish in Appendix~\ref{app:las} the linear scaling factor of valence bond fluctuations
in different phases:
  \begin{gather}
    \mathcal{F}_{AB} = {\alpha}_{\mathcal{F}} L + \mathcal{O}(\ln L).
    \label{eq:hc_fk}
  \end{gather} {In the gapless phase, we obtain
    ${\alpha}_{\mathcal{F}} = 3.84\tilde{c}_1$ where $\tilde{c}_1$ denotes
    the constant coefficient in the bond correlator (\ref{eq:i_func}) for a given set of $J_a$'s.
    In a gapped phase, we obtain
    ${\alpha}_{\mathcal{F}} \propto \xi^3$ with $\xi$ the
    correlation length.}  This approach then implies that with a rapidly growing
    correlation length, 
  ${\alpha}_{\mathcal{F}}$ must reach a maximum when undergoing a
  quantum phase transition from a gapped phase into the gapless
  intermediate regime (see Fig.~\ref{fig:2d}, a). 
  
  Numerically, we check these results by the method of finite-size scaling 
which starts from the anisotropic form (\ref{eq:bb_k}) of the function $I(i,j)$.
  The exact anisotropic   
  numerical calculations agree well with our previous
  $Y$-isotropic form approximation.
   In Fig.~\ref{fig:2d0}
  (b), we recover the linear scaling of $\mathcal{F}_{AB}$ in
  the gapless phase ($J_x = J_y = J_z$). The inset shows the scaling
  of $\mathcal{F}_A$, where the leading-order $L^2$ term ($0.391$) is
  dominated by the on-site bond fluctuations ($0.362$, from Eq.~(\ref{eq:osc})). Since 
  these on-site contributions are later subtracted,  $\mathcal{F}_{AB}$  contains more information about the entanglement properties.

%---------------------------------------------------------------------------------------------------------
  \subsubsection{Peak structure in linear scaling factor}
Shown in Fig.~\ref{fig:2d} (a), we continue our study by extracting numerically the linear scaling factor of valence bond fluctuations from different regimes of the phase diagram.
A peak structure centered at the quantum phase transition line is observed.
While the gapped region can be understood from the simultaneous evolution with the correlation length ($\alpha_{\mathcal{F}} \propto \xi^3$),
we check that the
    anisotropy effects in the $Y$-function are responsible for the
    decrease of $\alpha_{\mathcal{F}}$ when the system goes deeper
    into the gapless phase.} 
    
 It is noted that after the double summation in $\mathcal{F}_{AB}$ (\ref{eq:fab}), the bond fluctuations around the boundary (or domain wall) between two subsystems become the major contribution. Therefore, we can focus on the  short-range behavior of the anisotropic factor $Y(\vec{r})$ in the bond correlator (\ref{eq:yfunc}) along the direction perpendicular to the boundary. Illustrated by Fig.~\ref{fig:2d} (b), at short distances,  $Y(\vec{r})$ reaches the maximum value when $J_z$ {evolves} to the phase transition line.
Consequently, the amplitude of the linear scaling ${\alpha}_\mathcal{F}$ in $\mathcal{F}_{AB}$ would {drop} when we {decrease} $J_z$ in the gapless phase.     
    
    Fig.~\ref{fig:2d} (c)
  also includes the development of the linear scaling factor ${\alpha}_{\mathcal{F}}$ with different magnetic strengths. The signature of the peak structure in ${\alpha}_{\mathcal{F}}$ across the
  phase transition line is robust against small fields
  {($\kappa \le 0.10$)}. By increasing $\kappa$, the gap is enlarged. The anisotropic effects of the $Y(\vec{r})$ function originally dominant in the gapless region become reduced significantly, thus making the cusp of $\alpha_F$ more smooth. Stronger magnetic field effects are discussed qualitatively in Sec.~\ref{sec:afs}. 

\begin{figure}[t]
\begin{center}
   \includegraphics[width=.225\textwidth]{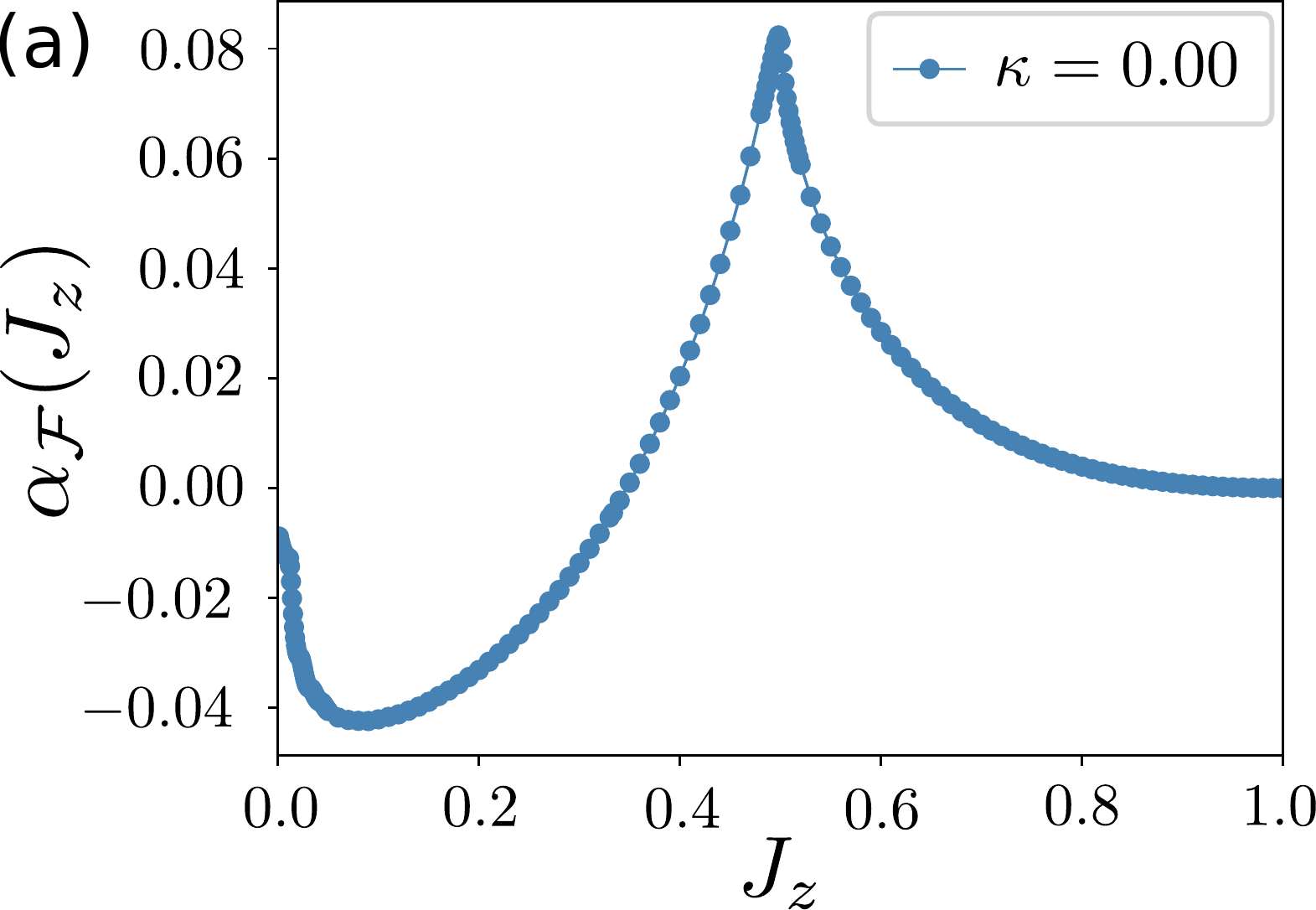} \ \includegraphics[width=.225\textwidth]{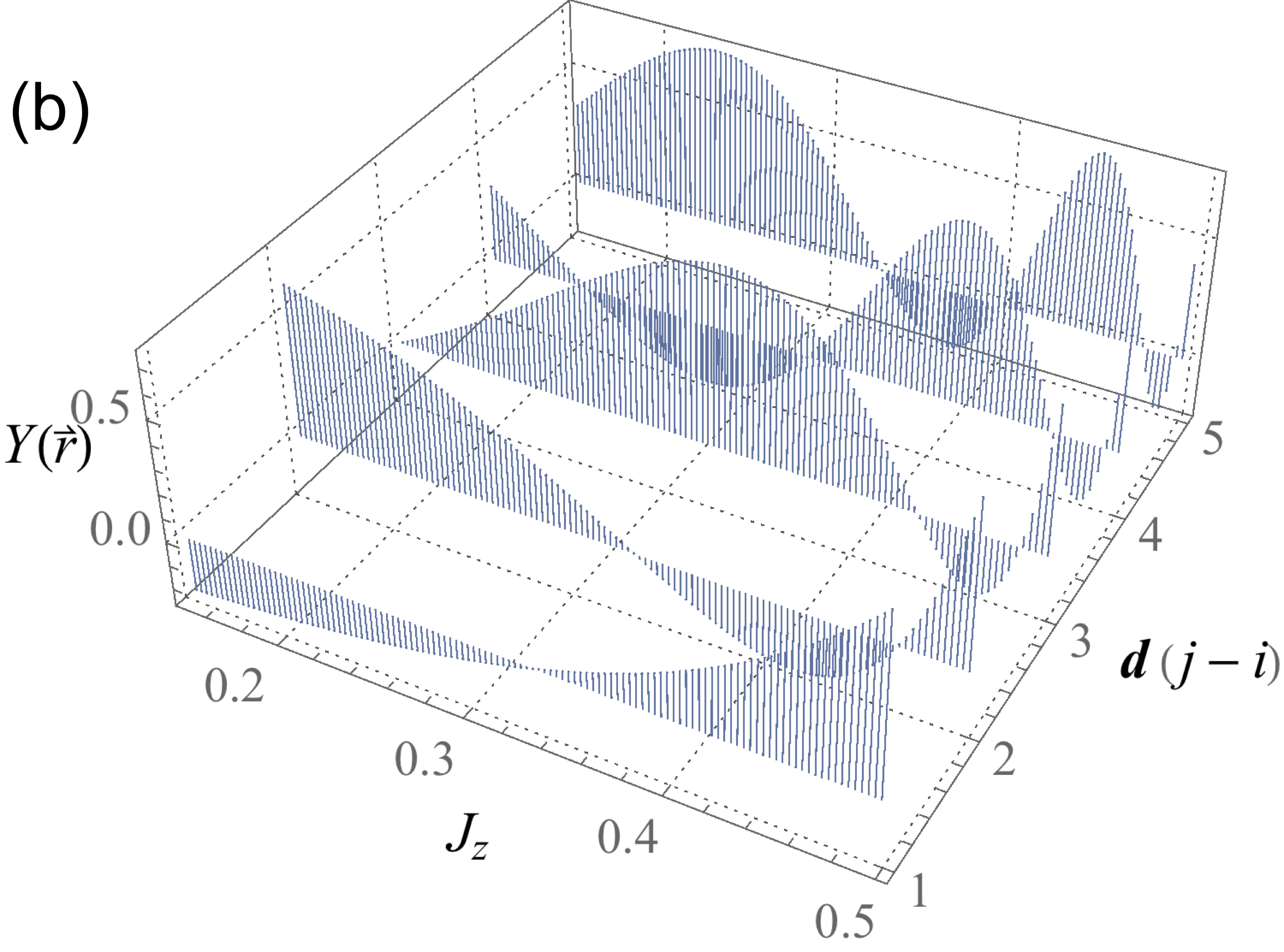} \\
   \vspace{0.25cm}
      \includegraphics[width=0.225\textwidth]{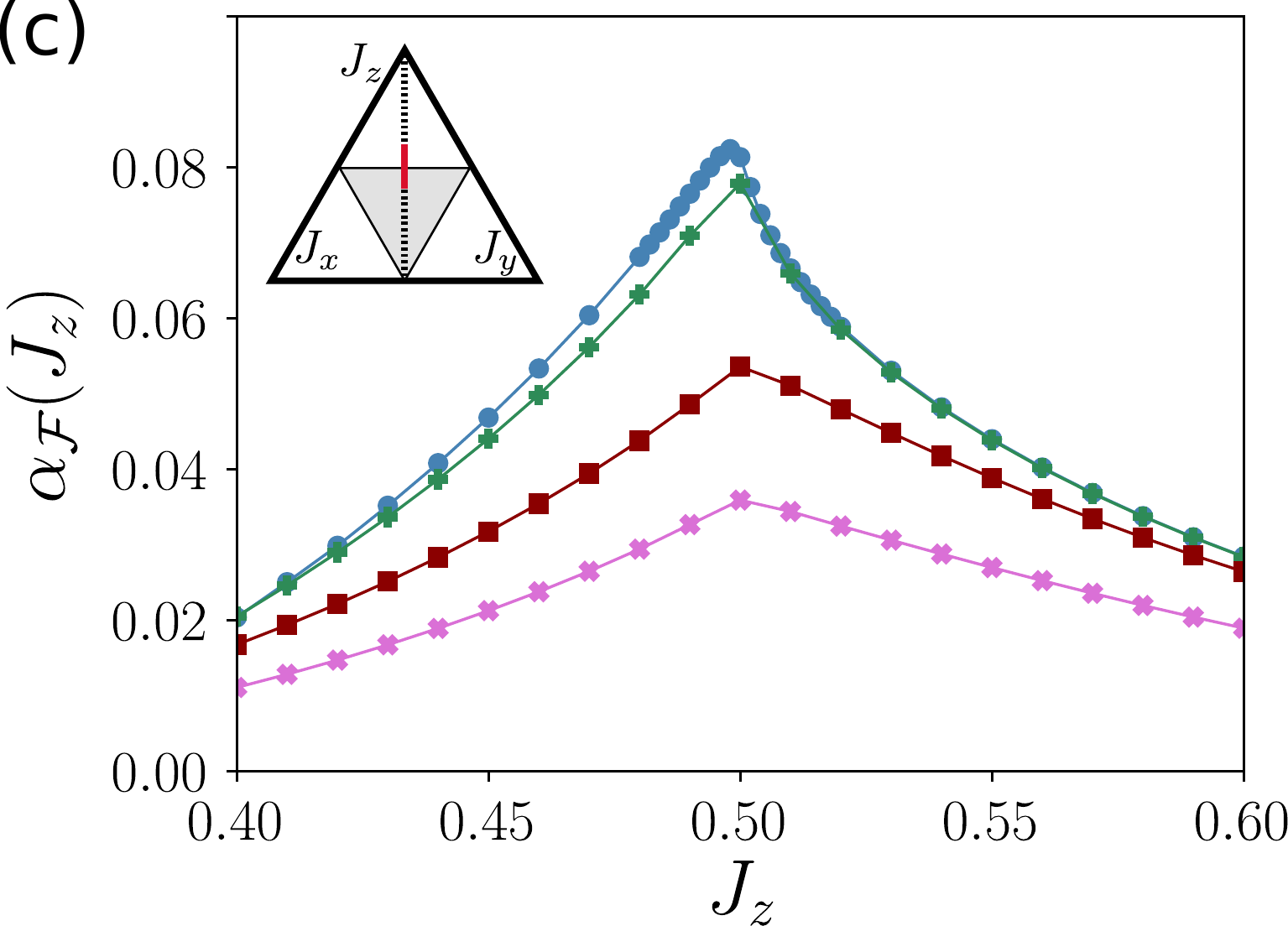} \includegraphics[width=0.225\textwidth]{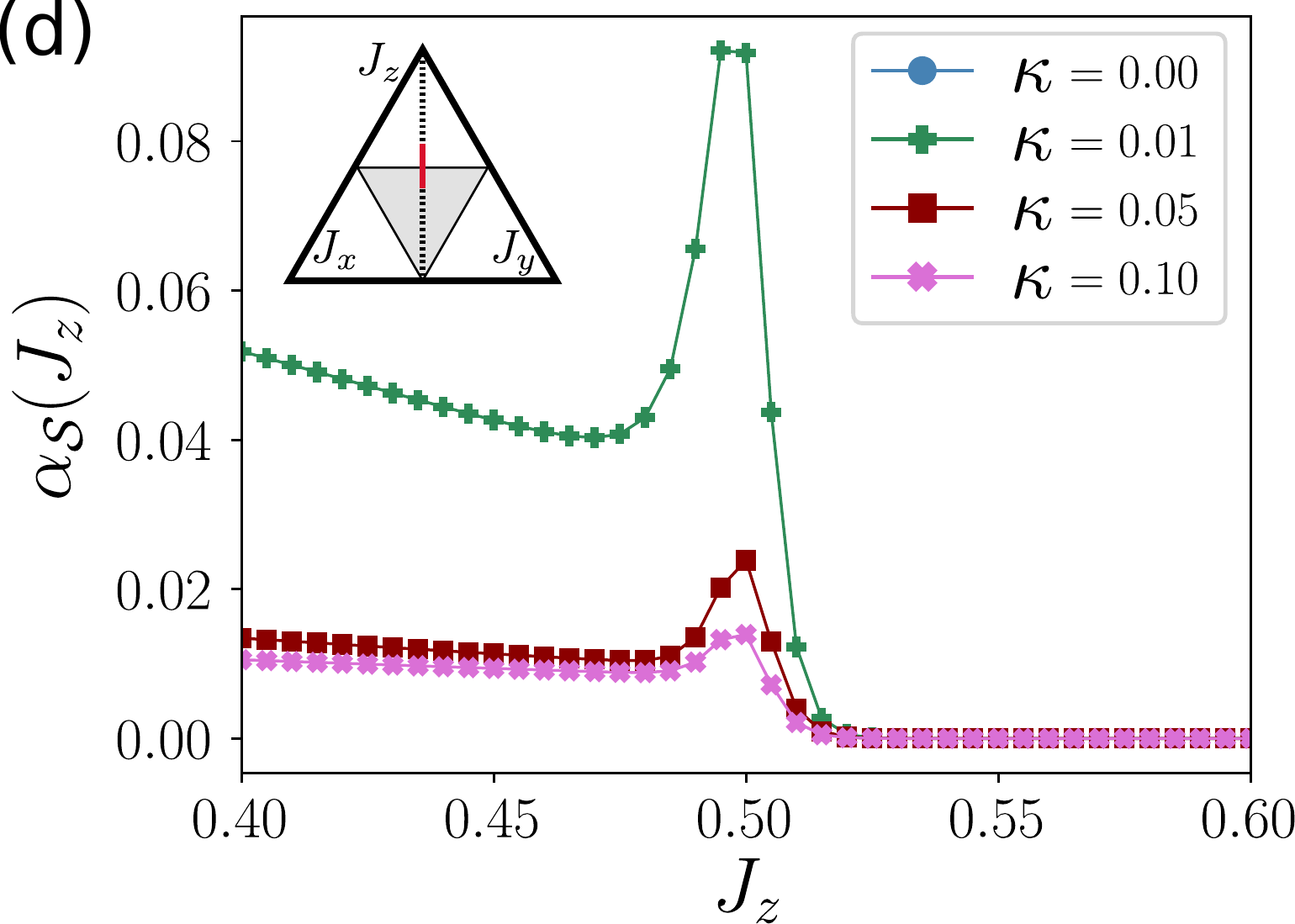}
        \end{center}
  \vskip -0.5cm \protect\caption[] {(a) Prefactor ${\alpha}_\mathcal{F}$ of the linear term in the bipartite fluctuations $\mathcal{F}_{AB}$ for the gapless intermediate phase ($J_z \le 0.5$) and the gapped phase with bonds polarized in the $z$ direction $(J_z > 0.5)$ of the Kitaev honeycomb model.  (b) Anisotropic function $Y(\vec{r})$ in the gapless phase. The relative vector is set along the direction of $\vec{n}_1$ {(perpendicular to the boundary between subsystems $A$ and $B$)}: $\vec{r} = \vec{r}_j -  \vec{r}_i = \textbf{d}(j-i)  = (j-i)\vec{n}_1$. 
  Effects of a small uniform magnetic field on the linear scaling factors of valence bond fluctuations and the Fermi entropy : (c) Prefactor
    ${\alpha}_\mathcal{F}$ in
    $\mathcal{F}_{AB}$; (d) Prefactor
    ${\alpha_\mathcal{S}}$ in
    $\mathcal{S}_{F}$.   In four plots,  the conventions for coupling constants $J_x = J_y$ and $J_x + J_y + J_z = 1$ are adopted.    
And the magnetic field strength 
    $\kappa$ varies in the range $[0.00, 0.10]$.    The calculation of $\mathcal{F}_{AB}$ is performed on a finite lattice cluster $\Omega = A \cup B = L \times L$ with the total length $L$ varying from $30$ to $100$.  For the finite-size scaling of $\mathcal{S}_F$, on the other hand,
    the length of the zigzag boundary across $x$-links is taken as
    $L=N_x \in [40, 100]$.}
    \label{fig:2d}
    \vskip -0.5cm
\end{figure} 
%------------------------------------------------------------------------------------------------
\subsubsection{Relation to Fermion entropy}

Now, we address the behavior of the entanglement entropy in the Kitaev
model.  As pointed out in Ref.~\cite{yao2010}, the total entanglement
entropy of the Kitaev honeycomb model consists of two pieces: the
gauge field part $\mathcal{S}_G = (L-1)\ln 2$ and the fermionic
contribution 
   \begin{gather}
     \mathcal{S}_F = {\alpha}_{\mathcal{S}} L + \mathcal{O}(1).
   \end{gather}
Since the
measurement of valence bond fluctuations preserves the $\mathbb{Z}_2$ gauge field
structure, $\mathcal{F}_{AB}$ probes the entanglement properties of the
fermion sector. Fortunately, $\mathcal{S}_F$ is responsible for all
the essential differences between the {Abelian and non-Abelian}
phases. 

We then extract the linear factor ${\alpha}_{\mathcal{S}}$
from $\mathcal{S}_F$ following the methods of Refs.~\cite{peschel2003,
  yao2010}. For completeness, we refer the readers to Appendix~\ref{app:entropy}
  where some details on the numerical approach to evaluate the Fermi entropy 
are presented.
It is found that in a small magnetic field $\kappa = 0.01$, the linear scaling factor
${\alpha}_{\mathcal{S}}$ shares the
same response as ${\alpha}_{\mathcal{F}}$ across the phase transition line shown in Fig.~\ref{fig:2d} (d).  
On top of that, once the magnetic field strength is increased, the peak structure of $\alpha_{\mathcal{S}}$ in
Fermi entropy disappears slightly more quickly  than the one in bond fluctuations.  In addition, it is relevant to observe that $\alpha_{\mathcal{S}}$ remains zero for $J_z>0.5$. 
Another interesting observation is that
as a function of $J_z$, the
magnitudes of $\alpha_{\mathcal{F}}$ and $\alpha_{\mathcal{S}}$ vary approximately in the same range $[0, 0.10]$.

To summarize, in two dimensions, we also find that valence bond fluctuations and the entanglement entropy of the Fermi sector
show the same area law scaling in all phases:
  \begin{gather}
    \mathcal{F}_{AB} \sim S_F. 
  \end{gather}
Moreover, their linear scaling factors act as signatures to characterize quantum phase transitions between the Abelian and non-Abelian
phases in the Kitaev honeycomb model.

%---------------------------------------------------------------------------------------------------------
\section{Concluding remarks}

In this final section, we would like to make brief remarks about the effects of stronger magnetic fields
on the gapless phase of the Kitaev honeycomb model. Two scenarios are presently discussed in the literature: one with excitations of flux pairs~\cite{janvsa2018, song2016}, and the other
with a  transition to another type of gapless spin liquid with U(1) symmetry~\cite{hickey2019}. We will propose
possible responses from the valence bond fluctuations respectively.

Afterwards, a comparison with the N{\'e}el state supported by antiferromagnetic Heisenberg interactions
will give us additional insights with regard to the application in quantum materials.

%----------------------------------------------------------------------------------------------------------
\subsection{Influence of Perturbations}
\label{sec:afs}
\subsubsection{Perturbed Kitaev QSLs with gauge-flux pairs}

For the Kitaev materials in a gapless spin liquid state, two types of gaps have been observed in the presence of a small tilted magnetic field~\cite{janvsa2018}
 \begin{gather}
   \Delta = \Delta_v + \Delta_f.
 \end{gather}
Here, $\Delta_v$ denotes the gap from the creation of a pair of fluxes (or visons) and $\Delta_f \propto h_xh_yh_z$ refers to the 
one induced by one matter Majorana fermion excitation. Our previous analysis of Sec.~\ref{sec:honeycomb} remains valid as long as $\Delta_v \gg \Delta_f$.

In Ref.~\cite{song2016}, it was suggested that one can construct an exact perturbed ground state with
even number of virtual fluxes by a unitary mapping $U$ from the unperturbed state $|\varphi_0\rangle$: $|\varphi \rangle = U | \varphi_0 \rangle$. The transformed spin operator 
takes the form~\cite{song2016}
 \begin{gather}
   U^\dagger \sigma_i^a U = iZ c_i c_i^a + \cdots + f_{ijk}^a i c_j c_k + \cdots, \label{eq:zf}
 \end{gather}
 with $Z = 1, f = 0 $ for the pure Kitaev model and $Z < 1, f \ne 0$ in the presence of perturbations. The nonzero $f$ parameters open a Majorana
 fermion gap instantaneously.
  
 One immediately notices for the valence bond operator $Q_i = \sigma_{i,1}^z \sigma_{i,2}^z$, 
 that the leading order contribution turns into
   \begin{gather}
      U^\dagger Q_i U = Z^2  (-ic_{i,1}c_{i,2}) + (\text{quartic terms}). \label{eq:tbc}
   \end{gather}
  Concerning the gauge Majorana fermions $c_i^a$, we have used the gauge convention form $u_{\langle i_1 i_2\rangle_z} = +1$ and all others being zero, acting on the unperturbed state.
  The first term in the transformed bond operator (\ref{eq:tbc}) has a rescaling factor $Z^2$. It leads to a $r^{-4}$ decay in valence bond correlation $I(i,j) = \langle Q_i Q_j \rangle_c$ 
  within the distance shorter than the correlation length ($r < \xi$).
  The second part contains  products of four matter Majorana fermions, which then result in much faster decays in bond correlations, for instance $r^{-6}$ and $r^{-8}$ over short distances.

Neglecting these higher order corrections arising from the $f$-decomposition and taking into account the general scaling rule on honeycomb lattice (\ref{eq:gsl0}), we establish that the valence bond fluctuations still show an area law
 \begin{gather}
   \mathcal{F}^{\text{VB}}_{AB, \text{perturbed}}  = \alpha_{\mathcal{F}, \text{perturbed}}\cdot L + \mathcal{O}(\ln L),
 \end{gather}
The linear scaling factor is rescaled according to 
  \begin{gather}
    \alpha_{\mathcal{F}, \text{perturbed}} \simeq Z^4 \alpha_{\mathcal{F},0}, \label{eq:alphap}
  \end{gather}
 where $\alpha_{\mathcal{F},0}$ denotes the prefactor of the linear term reminiscent of  the zero-flux Kitaev spin liquids. Based on Sec.~\ref{sec:bfha}, we thus find
  \begin{gather}
    \alpha_{\mathcal{F},0} \propto \xi_f^3 \propto \Delta_f^{-3}.
  \end{gather} 
With excitations of flux pairs, the linear scaling factor
 now combines two pieces of information: the vison gap $\Delta_v$ determining the amplitude of $Z^4$ and the Majorana fermion
 gap $\Delta_f$ coming into play through $\alpha_{\mathcal{F},0}$.
 
 It may be relevant to mention that the two-spin fluctuations become already nonzero in the perturbed limit. From Eq.~(\ref{eq:zf}), one gets contributions from the $f$-sector of Majorana fermions. Accompanied by an exponential decay in the spin-spin correlation,  we obtain
  \begin{gather}
   \mathcal{F}^{\text{TS}}_{AB, \text{perturbed}}  = \alpha^{\text{TS}}_{\mathcal{F}}\cdot L + \mathcal{O}(\ln L),
 \end{gather}
 where
  \begin{gather}
   \alpha^{\text{TS}}_{\mathcal{F}} \propto (f^z)^2 \cdot \Delta_f^{-3}.
 \end{gather}
When $\Delta_v \gg \Delta_f$, no excitation of fluxes is allowed. When $f^z \to 0$,  $\alpha^{\text{TS}}_{\mathcal{F}} \to 0$, we check the result of vanishing two-spin fluctuations in the solvable limit (\ref{eq:tsk}).

%----------------------------------------------------------------------------------------------------------
\subsubsection{Transition to U(1) gapless spin liquids}

If one continues to increase the strength of the uniform magnetic field, from numerical simulations~\cite{hickey2019},
while the Kitaev ferromagnet produces a trivial polarized phase (PL), the Kitaev antiferromagnet might give rise
to an intermediate phase supporting U(1) gapless spin liquids (GSL).

In the PL phase, there is no correlation between two subsystems and both the two-spin and valence bond fluctuations vanish
 \begin{gather}
   \mathcal{F}^{\text{TS/VB}}_{AB, \text{PL}}  = 0.
 \end{gather}

For the GSL phase in the Kitaev antiferromagnet, one can assume a gapless spinon Fermi surface coupled to a U(1)
gauge field. In an effective picture of complex Abrikosov fermions~\cite{hickey2019}, a spin operator is mapped onto the product of fermions 
   $2\vec{S} =f_\alpha^\dagger \vec{\sigma}_{\alpha \beta} f_\beta$, with spin index $\alpha, \beta = \uparrow, \downarrow$
   and a U(1) symmetry $f_\alpha^\dagger \to e^{i\theta} f_\alpha^\dagger$.
From this perspective, the spin and bond correlations follow power-law decays ($r^{-4}$, $r^{-8}$ respectively) in the gapless phase.
We predict that an ``enhancement" might be observed in the prefactor of the area-law bipartitie fluctuations
 \begin{gather}
   \mathcal{F}^{\text{TS/VB}}_{AB, \text{GSL}}  = \alpha^{\text{TS/VB}}_{\mathcal{F}, \text{GSL}}\cdot L + \mathcal{O}(\ln L).
 \end{gather}
The existence of a similar peak structure in $\alpha_{\mathcal{F}}$ between the gapped Kitaev spin liquids and U(1) gapless spin liquids is possible and can be tested numerically in the future.

%--------------------------------------------------------------------------------------------
\subsection{Comparison with the N{\'e}el phase}
\label{sec:neel}
In the end, to make a closer link with quantum materials, it is perhaps useful to compare
the 
obtained behavior of bond-bond
correlation functions from the ones of the two-dimensional Heisenberg
model, {\it i.e.}, of a N\' eel ordered phase subject to spin-wave
excitations.  When antiferromagnetic Heisenberg interactions are
dominant, the modified spin-wave theory predicts that a staggered
magnetic field is required to stabilize the N{\'e}el state at zero
temperature for finite lattices~\cite{takahashi1989, song2011}.

Performing a spin-wave analysis in Appendix~\ref{app:sw}, then we find that the same valence
bond correlation shows: 
  \begin{gather}
      I(i,j) = c_0 + {c_1}{r}^{-1}
   \end{gather}
   {with
  $c_0 = 0.131$ and $c_1 = 0.141$}. 
 
  As a result, the bipartite
fluctuations now follow a volume square law:
  \begin{gather}
    \mathcal{F}_{AB} \propto L^4,
  \end{gather}
  arising from the non-vanishing
  long-range correlation of $c_0$. Measuring the precise {leading
  order scalings} then allows to probe the phase, Kitaev spin liquid
versus N\' eel state, of a two-dimensional quantum material. We
emphasize here that the entanglement entropy of the N\' eel state
still reveals an area law~\cite{song2011}, as in the Kitaev spin
model. The violation of the lower bound (\ref{eq:lb}) originates from
the finite-size regularization procedure taken in the modified spin-wave theory.

%-----------------------------------------------------------------------------------------------
\subsection{Conclusion}

To summarize, we have found a general relation between the valence
bond fluctuations and the entanglement entropy of the 
Kitaev spin model in one and two dimensions. Valence bond fluctuations
appear as a relevant tool to identify phases and phase transitions of Majorana magnetic quantum systems. 
Application to three-dimensional
systems~\cite{Maria,Mariareview} can be studied in the future.
\\

%-----------------------------------------------------------------------------------------------------------
\section*{Acknowledgements}

 We would like to thank two anonymous referees for their valuable suggestions on 
  comparing different types of spin liquids and 
 on  developing the non-perturbative
 regime.
 This work has also benefitted a lot from discussions
with L. Henriet, L. Herviou, N. Laflorencie, C. Mora, F. Pollmann,
S. Rachel, G. Roux, H.-F. Song, A. Soret, at the DFG meetings FOR 2414
in Frankfurt and G{\"o}ttingen, at CIFAR meetings in Canada and at the
conference in Montreal related to the workshop on entanglement,
integrability, topology in many-body quantum systems. Support by the
Deutsche Forschungsgemeinschaft via DFG FOR 2414 is acknowledged as
well as from the ANR BOCA. KP acknowledges the support by the Georg
H. Endress foundation. Numerical calculations performed using the
ITensor C++ library, http://itensor.org/.

%---------------------------------------------------------------------------------------------------------------------------------------------------
\renewcommand{\theequation}{\Alph{section}\arabic{equation}}
\appendix

%-----------------------------------------------------------------------------------------
\section{SU(2) quantum spin systems}

Here, we evaluate the two-spin fluctuations and valence bond fluctuations for the SU(2)-symmetric quantum spin models both in one and two dimensions.
We differentiate two cases: the one where the singlets resonate with equal weights and the other where they decay exponentially over distance.

\label{app:su2}
\subsection{Finite-size singlet state}
\begin{figure}[t]
   \begin{center}
		\includegraphics[width=0.28\linewidth]{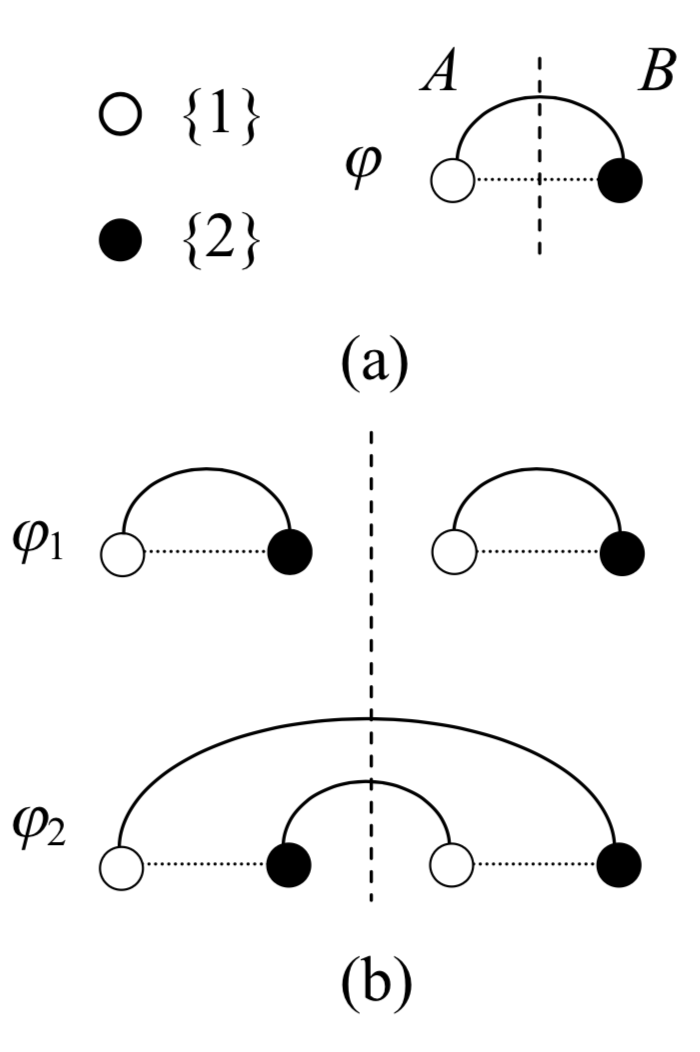}  \includegraphics[width=0.8\linewidth]{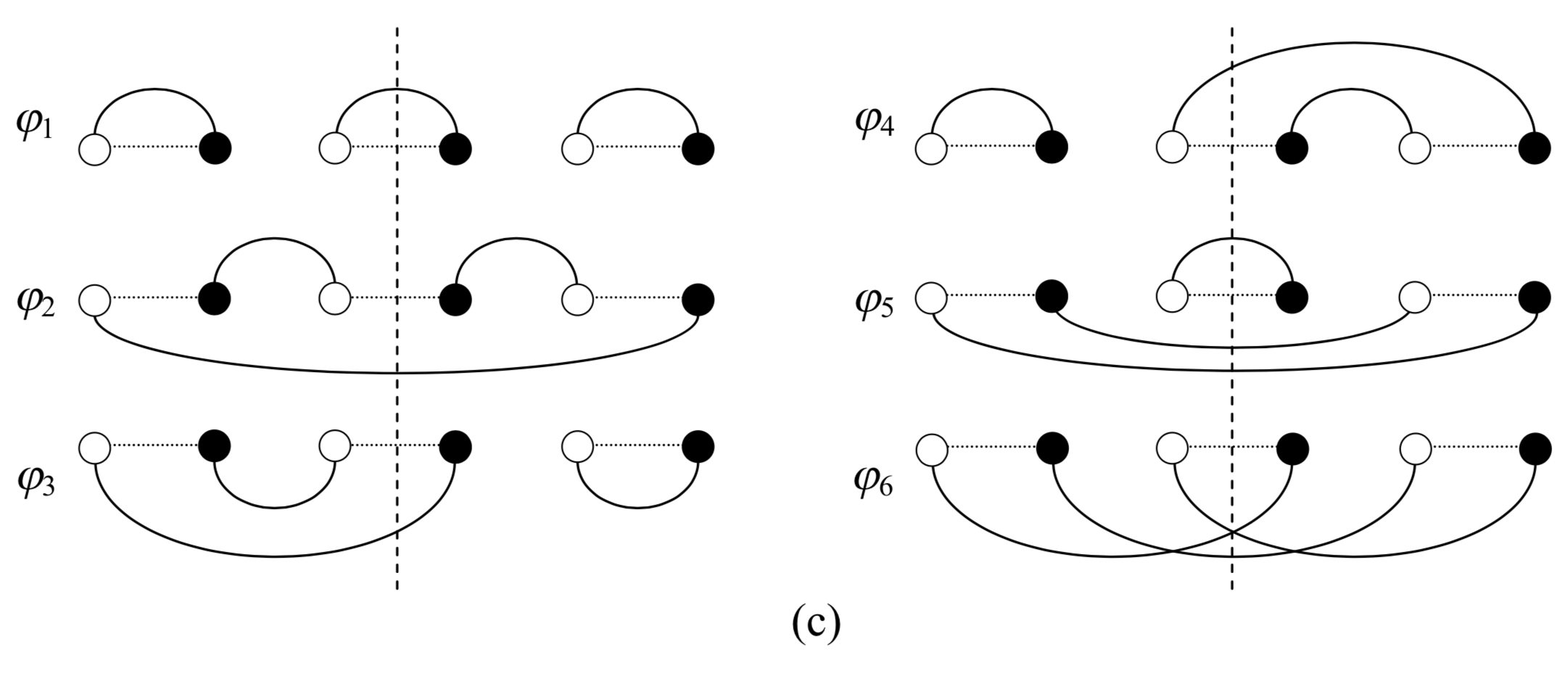} 
    \end{center}
		  \vskip -0.5cm \protect\caption[]
     {Possible pairing configurations for the $N$-site singlet state: (a) $N = 2$; (b) $N = 4$; (c) $N = 6$. We allow for entanglement on different sublattices. Two sites paired by a solid curve form a local valence bond state: $|\frown\  \rangle_{\overline{(i,1)(j,2)}} =  ( | \uparrow_{i,1} \rangle | \downarrow_{j,2} \rangle - | \downarrow_{i,1} \rangle | \uparrow_{j,2} \rangle)/\sqrt{2}$. }
	\label{fig:singlet}
	\vskip -0.5cm
\end{figure} 
We start from one dimension. Fig.~\ref{fig:singlet}  shows all the pairing states $\{ |\varphi\rangle_i \}$ for $N = 2, 4, 6$-site singlets. $N$ can also be interpreted as the maximum resonating range of the VB states on an infinite chain. 
As $N \to \infty$, the system approaches the critical point. 

When $N = 2$, we have a unique singlet state $|\Phi_0\rangle = |\varphi\rangle$ shown in Fig.~\ref{fig:singlet} (a). Once the boundary between two subsystems $A$ and $B$ is set on the bond, one recovers immediately $\mathcal{F}^{\text{TS}} = 1$, $ \mathcal{F}^{\text{VB}} = 0$, $\mathcal{S}^{\text{VB}} = \ln 2$ and the relation $\mathcal{S}^{\text{VB}}/\mathcal{F}^{\text{TS}} = \ln 2$. Nevertheless, if the boundary is off the bond, $\mathcal{F}^{\text{TS}} = \mathcal{F}^{\text{VB}} =\mathcal{S}^{\text{VB}} = 0$. In the following analysis, we always keep the size of the two subsystems to be the same. 

For $N= 4$ in Fig.~\ref{fig:singlet} (b),  a simple normalized singlet state reads
    $|\Phi_0\rangle =  (| \varphi_1 \rangle +|  \varphi _2 \rangle)/\sqrt{3}$ with an overlap between two pairings $\langle \varphi_1 | \varphi_2 \rangle = 1/2$. Also, we have assumed  the system has equal chance to stay in each pairing state. Then
  \begin{gather}
    \mathcal{S}^{\text{VB}} = \frac{0 + 2\ln 2 /\sqrt{3}}{2/\sqrt{3}}  = \ln 2, \notag \\
     \mathcal{F}^{\text{TS}} = 0 + \left(\frac{1}{\sqrt{3}}\right)^2 \cdot 2 = \frac{2}{3}, \quad \mathcal{F}^{\text{VB}} = 0 + \left(\frac{1}{\sqrt{3}}\right)^2= \frac{1}{3}.
    \end{gather}
We find a new relation 
  \begin{gather}
    \frac{\mathcal{S}^{\text{VB}}}{\mathcal{F}^{\text{TS}} + \mathcal{F}^{\text{VB}}}= \ln 2. \label{eq:ln}
  \end{gather} 
  
  Yet Eq.~(\ref{eq:ln}) does not hold if $N > 4$.
   We can check the case $N = 6$ shown in Fig.~\ref{fig:singlet} (c). 
 The system allows $(N/2)! = 6$ pairing configurations and the normalization factor $\mathcal{D}$ becomes larger than 6 on account of the overlaps:
   \begin{gather}
     |\Phi_0\rangle = \frac{1}{\sqrt{\mathcal{D}}} \sum_{p=1}^{n_p=6} |\varphi _p \rangle,  \quad \mathcal{D} = \sum_{p,p'=1}^{n_p=6} \langle \varphi_p | \varphi_{p'} \rangle  = 18.
   \end{gather}
While each $|\varphi_{p} \rangle$ is responsible for $\mathcal{S}^{\text{VB}}$ and $\mathcal{F}^{\text{TS}}$, 
 only $|\varphi_{p=4,5}\rangle$ contribute to $\mathcal{F}^{\text{VB}}$. 
  \begin{gather}
   \mathcal{S}^{\text{VB}} = \frac{10 \cdot (\ln 2) / \sqrt{\mathcal{D}}}{6/\sqrt{\mathcal{D}}} = \frac{5 \ln 2}{3}, \notag \\
   \mathcal{F}^{\text{TS}} = \left( \frac{1}{\sqrt{\mathcal{D}}} \right)^2   \cdot 10 = \frac{5}{9}, 
   \quad \mathcal{F}^{\text{VB}} = \left( \frac{1}{\sqrt{\mathcal{D}}} \right)^2  \cdot 2 = \frac{1}{9}.
 \end{gather}
  Therefore, ${\mathcal{S}^{\text{VB}}}/({\mathcal{F}^{\text{TS}} + \mathcal{F}^{\text{VB}}})  = ({5}/{2}) \ln 2 > \ln 2$.
  
In two dimensions, for a gapped VB state composed of $N$-site singlets, the generalized result becomes
      \begin{gather}
        \mathcal{S}^{\text{VB}} \propto \ln 2 \cdot n, \quad  \frac{\mathcal{S}^{\text{VB}}}{\mathcal{F}^{\text{TS}} + \mathcal{F}^{\text{VB}}} \ge \ln 2.
      \end{gather}
$n$ denotes the number of singlets that the boundary crosses. Both relations above take the equality $``="$ if the resonating range $N \le 4$. Meanwhile, we recover the area law of bipartite fluctuations
   \begin{gather}
     \mathcal{F}^{\text{TS}} = f^{\text{TS}}(N) \cdot n, \quad \mathcal{F}^{\text{VB}} = f^{\text{VB}}(N) \cdot n,
    \end{gather}
where the coefficients depend on the size of the singlets. For instance, we have $f^{\text{TS}}(2) = 1, f^{\text{VB}}(2) = 0; f^{\text{TS}}(4) = 2/3, f^{\text{VB}}(4) = 1/3;  f^{\text{TS}}(6) = 5/9, f^{\text{VB}}(6) = 1/9$.

%---------------------------------------------------------------------------------------
\subsection{General singlet state}

Next, we consider a gapped system where the singlet bonds decay exponentially  with distance, and $\mathcal{F}^{\text{TS}}$ and $\mathcal{F}^{\text{VB}}$ still show an area law as the entanglement entropy.

We first test the 1D case. A general VB state can be constructed as
  \begin{align}
    |\Phi_0 \rangle = \frac{1}{\sqrt{\mathcal{D}}} \sum_{p}  \prod_{\overline{(i,1)(j,2)} \in p}  
         \frac{e^{-|i-j|/\xi} }{\sqrt{2}} ( | \uparrow_{i1} \rangle | \downarrow_{j2} \rangle - | \downarrow_{i1} \rangle | \uparrow_{j2} \rangle)
  \end{align}
where $\xi$ stands for the correlation length of dimers. The probability of two sites $(i,1)$ and $(j,2)$ being paired reads
 \begin{gather}
    P(\overline{(i,1)(j,2)}) = P(|i-j|) = c e^{-2|i-j|/\xi}.
   \end{gather}
Correspondingly, on a chain with a total length $2L$ ($1 \ll \xi \ll L$),
   \begin{align}
    \mathcal{F}^{\text{TS}}_{\text{1D}} 
       &= \sum_{(i,\alpha) \in A} \sum_{(j, \beta) \in B} P(|i-j|)  \notag \\
       &= \sum_{i=1}^{L/2} \sum_{j=L/2+1}^{L} P(|i-j|) \cdot 2 = \frac{c}{2} \xi^2 + \mathcal{O}(\xi), \\
     \mathcal{F}^{\text{VB}}_{\text{1D}} 
       &= \sum_{i=1}^{L/2} \sum_{j=L/2+1}^{L} P(|i-j|)^2 = \frac{c^2}{16} \xi^2 + \mathcal{O}(\xi).
  \end{align}
$\mathcal{F}^{\text{TS/VB}}_{\text{1D}}$ becomes a constant proportional to the square of the correlation length.

In two dimensions, one can apply the general scaling rule of Appendix~\ref{app:gsr} on $\mathcal{F}^{\text{TS/VB}} = \sum_{i \in A, j \in B} I(i,j)$. Since the correlators $I(i,j)$ follow an exponential decay  (as in one dimension), the bipartite fluctuations between two subsystems should be proportional to the perimeter $x$ of the boundary
  \begin{gather}
    \mathcal{F}^{\text{TS/VB}}_{\text{2D}}  = b^{\text{TS/VB}}  \cdot x + \mathcal{O}(x).
      \end{gather}
Hence, we find a similar area law in the two types of fluctuations as for the VB entanglement entropy in (\ref{eq:vb}).

%-----------------------------------------------------------------------------------------
\section{Kitaev spin chain}
\label{app:lk1}
{Here, we give a mathematical proof of Eq.~(\ref{eq:fc}). }

%------------------------------------------------------------------------------------
\subsection{Quantum critical point $J_1 = J_2$}
\label{sec:chain_critical}
For the critical Kitaev spin chain at the gapless point $J_1 = J_2$, we first evaluate the bipartite fluctuation within subregion $A$:
  \begin{gather}
   \mathcal{F}_A = \sum_{i,j\in A} I(i,j) = \sum_{i,j \in A} I(\left| i - j\right|).
 \end{gather}  
The bond correlation depends only on the difference of two variables. One can thus convert the double sum into a single sum through
  \begin{gather}
     \sum_{i,j \in A} I(\left| i - j\right|) = I(0)l_A + 2\sum_{k=1}^{l_A} (l_A - k)I(k).
     \label{eq:ds0}
  \end{gather}
From the expression (\ref{eq:ik1}) of $I(k)$: $1/[{\pi^2} (k^2 - 1/4)]$ for $k \ne 0$; $1-4/\pi^2$ for $k = 0$, we get
  \begin{align}
     \sum_{k=1}^{l_A} I(k) &=  \frac{4}{\pi^2} \sum_{k=1}^{l_A}\frac{1}{4k^2-1}  \notag \\
     &= \frac{2}{\pi^2} \left( 1- \frac{1}{2l_A+1}\right) =  \frac{2}{\pi^2} +    \mathcal{O}(l_A^{-1}), \notag \\
     \sum_{k=1}^{l_A} k I(k) &= \frac{1}{\pi^2} \sum_{k=1}^{l_A} 2k \left( \frac{1}{2k-1} - \frac{1}{2k+1} \right) \notag \\
     &= \frac{1}{\pi^2} \sum_{k=1}^{2l_A} \left( \frac{k+1}{k} - \frac{k-1}{k} \right)    \sigma_k - \frac{l_A}{2l_A + 1} \notag \\
     &= \frac{2}{\pi^2} \sum_{k=1}^{2l_A} \frac{\sigma_k}{k} - \frac{l_A}{2l_A + 1},
        \end{align}
  with $\sigma_k = 1$ for $k = \text{odd}$ and $\sigma_k = 0$ for $k = \text{even}$.
It is convenient to re-express the finite sum in terms of the difference of two infinite sums
    \begin{align}
      \sum_{k=1}^{2l_A} \frac{\sigma_k}{k} &= \sum^{\infty}_{k=1} \frac{\sigma_k}{k} -  \frac{\sigma_{k+2l_A}}{k+2l_A}  \notag \\
      &= \sum_{k=1}^{\infty} \left( \frac{1}{k} - \frac{1}{2k} \right) - \left( \frac{1}{k+2l_A} - \frac{1}{2k+2l_A} \right).
    \end{align}
 In the second equality, we have also used the relation: $(\text{odd terms}) = (\text{all terms}) - (\text{even terms})$.
 Now we can apply the properties of the digamma function, which shares the series representation related to the Euler's constant $\gamma \simeq 0.57721$, as well as the asymptotic expansion   
 \begin{align}
     &\psi (x) = - \gamma + \sum_{k=1}^{\infty} \left( \frac{1}{k} - \frac{1}{k+x-1} \right),  \notag \\
       &\psi (x \to \infty) = \ln x - \frac{1}{2x} - \frac{1}{12x^2} + \mathcal{O} (x^{-4}).
   \end{align}
 Therefore,
   \begin{align}
     \sum_{k=1}^{2l_A} \frac{\sigma_k}{k} 
         &= \psi (2l_A + 1) + \gamma - \frac{1}{2} \left( \psi(l_A + 1) + \gamma \right) \notag \\
         &= \frac{1}{2}\ln l_A + \frac{\gamma}{2} + \ln 2 + \mathcal{O}(l_A^{-1}).
    \end{align}
For a bipartition $l_A = l_B = L/2$, we then get the critical scaling of $\mathcal{F}_A$  and at the same time, $\mathcal{F}_{AB}$ from their relation (\ref{eq:ffab})
    \begin{align}
    \mathcal{F}_A &= l_A - \frac{2}{\pi^2} \ln l_A - \frac{2}{\pi^2} \left( \gamma + 2\ln 2 - \frac{1}{2}\right) + \mathcal{O}(l_A^{-1}), \notag \\
    \mathcal{F}_{AB} &= \frac{1}{\pi^2} \ln l_A + \frac{1}{\pi^2} \left( \gamma + \ln 2 - \frac{1}{2}\right) + \mathcal{O}(l_A^{-1}).
  \end{align}    
  
  %-------------------------------------------------------------------------------------------------------------------------------------------
\subsection{Gapped regime $|J_1| > |J_2|$}
 \label{sec:chain_gapped}
 
We continue to study the gapped phase of the Kitaev spin chain  with  negative exchanges such that $|J_1| > |J_2|$, such that the strong bonds occur on the {$x$-links}. In Eqs.~(\ref{eq:kci1}) and (\ref{eq:kci2}), we predict that the bond correlator behaves as
$I(i,j) = c_1 | i-j |^{-2}$ for $| i-j | \le \xi$ and
$I(i,j) = c_2 e^{- | i-j |/\xi }$ for $| i-j | > \xi$, with a correlation length $\xi \sim \left| J_2 - J_1 \right|^{-1}$. For $1<\xi< l_A = l_B = L/2$, the valence bond fluctuations between two subregions become
\begin{align}
   \mathcal{F}_{AB} &= \sum_{i=1}^{L/2} \sum_{j=L/2+1}^{L} I(i,j)  \notag \\
     &= \sum_{k=1}^{\xi} k \frac{c_1}{k^2} + \sum_{k = \xi+1}^{L/2} k c_2 e^{- \frac{k}{\xi} } + \sum_{k=L/2 + 1}^{L} \left( L-k \right) c_2 e^{- \frac{k}{\xi} }.
\end{align}
Approximating the single summation by an integral and supposing
$\xi \ll L$, one obtains
  \begin{gather}
     \mathcal{F}_{AB} = c_1 \ln \xi + c_2 e^{-1} \left( 2 \xi^2 - \xi \right) + \mathcal{O}\left( 1  \right).
  \end{gather}

%------------------------------------------------------------------------------------------------
\section{Asymptotic form of bond correlator in the intermediate gapless phase}
\label{app:kitaev}
Here, we derive the power-law behavior of bond correlation functions in the intermediate gapless spin liquid phase of the Kitaev honeycomb model. We assume a simple case where three Ising couplings share the same strength: $J_x = J_y = J_z = J$. 

When $\kappa = 0$, the valence bond correlator (\ref{eq:bb_k}) can be re-expressed as the product of two sums
  \begin{align}
    I(i,j) &= - \langle -ic_{i, 1}c_{j, 2} \rangle  \langle -ic_{j, 1}c_{i, 2} \rangle, \notag \\
     \langle -ic_{i, 1}c_{j, 2} \rangle &= \frac{1}{N} \sum_{\vec{k}} e^{i\vec{k} \cdot (\vec{r}_j - \vec{r}_i)} \frac{f(k)^*}{|f(k)|}, \notag \\
      \langle -ic_{j, 1}c_{i, 2} \rangle &= \frac{1}{N} \sum_{\vec{q}} e^{-i\vec{q} \cdot (\vec{r}_j - \vec{r}_i)} \frac{f(q)^*}{|f(q)|}. 
      \label{eq:iks}
  \end{align}
The main contribution comes from the two Dirac points $\pm \vec{k}^* = \pm (k_x^*, k_y^*)= \pm ({4\pi}/{3}, 0)$ which satisfy $|f(\pm \vec{k}^*)| = 0$. It allows us to approximate the summation by an expansion around each Dirac point within a small radius $\xi$: $\vec{k} \in \Omega(\pm \vec{k}^*, \xi)$. 

For the first sum in Eq.~(\ref{eq:iks}), around one Dirac point $ \vec{k}^* = (4\pi/3, 0)$ we get
  \begin{gather}
    \frac{f(\vec{k})^*}{|f(\vec{k})|} = -\cos \theta' + i\sin \theta'.
  \end{gather}
 Here $\theta'$ is the angle between the relative vector around the Dirac cone $\delta \vec{k} = \vec{k} - \vec{k}^*$ and the $x$ axis. It is clear to see that $I(i, j)$ is anisotropic. To simplify the exponential, we denote the direction of the two unit cells  as $\vec{r} = \vec{r}_{j} - \vec{r}_i = (r\cos{\theta^*}, r\sin{\theta^*})$ with $\theta^*$ the angle between vectors $\vec{r}$ and $\vec{k}^*$. Then,
    \begin{gather}
     e^{i\vec{k}\cdot(\vec{r}_{j} - \vec{r}_i)} = e^{i\vec{k}^*\cdot \vec{r}} e^{i\delta \vec{k}\cdot \vec{r}} =  e^{i\vec{k}^*\cdot \vec{r}} e^{i{\delta k}  r \cos \theta},
    \end{gather}
 with $\theta$ the relative angle between $\delta \vec{k}$ and $\vec{r}_{j} - \vec{r}_i$. 
 
 Now we can evaluate the summation by taking the continuum limit
  \begin{align}
     \langle -ic_{i, 1}c_{j, 2} \rangle_{\vec{k}^*}
     = & \frac{(-1)}{(2\pi)^2} \cdot \frac{\sqrt{3}}{2} \int_{0}^{\xi} k dk \int_{0}^{2\pi} e^{ikr\cos{\theta}} \notag \\
      & \times e^{i(\vec{k}^*\cdot \vec{r}-\theta^*)}(\cos \theta + i \sin \theta) d\theta.
    \label{eq:j1}
  \end{align}
  The factor $2$ comes from the contribution of two Dirac points.  The other factor $\sqrt{3}/2$ originates from a change of basis from $dk_1 dk_2$ in the Brillouin zone (with unit vectors $\vec{n}_1$ and $\vec{n}_2$) to $dk_x dk_y$. We have also used the relation $\theta + \theta' = \theta^*$. 
  
  Via a change of variables  $k' = kr$, one reaches
  \begin{align}
    \langle -ic_{i, 1}c_{j, 2} \rangle_{\vec{k}^*} 
        &= - \frac{\sqrt{3}}{8\pi^2}e^{i(\vec{k}^*\cdot \vec{r}-\theta^*)}\int_{0}^{\Lambda} \frac{kdk}{r^2} \int_{0}^{2\pi} \cos{\theta} e^{ik\cos{\theta}} d\theta \notag \\
        &= -\frac{t(\Lambda)}{2 r^2} i e^{i(\vec{k}^*\cdot \vec{r}-\theta^*)},
  \end{align}
  where
 $t(\Lambda) = \sqrt{3}/(2\pi)\cdot \int_{0}^{\Lambda} J_1(k)kdk$ with a cutoff $\Lambda = \xi r$ and inside the integral $J_1(k)$ denotes the Bessel function of the first kind.
 
A similar expansion around the other Dirac point $-\vec{k}^* = (-4\pi/3, 0)$ would give an additional phase factor $e^{i(-\vec{k}^*\cdot \vec{r}-(\pi-\theta^*))}$. 
Thus, the total contribution reads
    \begin{align}
       \langle -ic_{i, 1}c_{j, 2} \rangle &=  \langle -ic_{i, 1}c_{j, 2} \rangle_{\vec{k}^*}  + \langle -ic_{i, 1}c_{j, 2} \rangle_{-\vec{k}^*}  \notag \\
                                                     &= \frac{t(\Lambda)}{r^2} \sin(\vec{k}^*\cdot \vec{r}-\theta^*). \label{eq:iks1}
  \end{align}
  
   For the second sum in Eq.~(\ref{eq:iks}), we only need to change $\vec{r}$ to $-\vec{r}$ and adjust the relative angle from $\theta^*$ to $\theta^* - \pi$:
  \begin{gather}
    \langle -ic_{j, 1}c_{i, 2} \rangle = \frac{t(\Lambda)}{r^2} \sin(\vec{k}^*\cdot \vec{r}+\theta^*). \label{eq:iks2}
   \end{gather}

Combining Eqs.~(\ref{eq:iks}), (\ref{eq:iks1}) and (\ref{eq:iks2}), we then recover the $r^{-4}$ scaling of the bond correlator in the gapless phase~\cite{shuo2008}:
  \begin{gather}
     I(i,j) = \frac{\widetilde{c}_1}{r^4}.
  \end{gather}
Furthermore, from our calculations the amplitude $\widetilde{c}_1$ retrieves an anisotropic factor $Y(\vec{r})$:
 \begin{gather}
      \widetilde{c}_1 = t^2(\Lambda) \cdot Y(\vec{r}), \notag \\
       Y( \vec{r}) = \cos^2(\vec{k}^*\cdot \vec{r}) - \cos^2(\theta^*).
      \label{eq:hi}
    \end{gather}   
One can also verify that the forms of $\tilde{c}_1$ and of the anisotropic $Y$-function in Eq.~(\ref{eq:hi}) are valid for the whole gapless region.  

%------------------------------------------------------------------------------------------------------------------------------------------------
\section{Bipartite fluctuations on honeycomb geometry}
\label{app:las}

We evaluate here the bipartite fluctuations on the honeycomb lattice, involving the lattice summation.
%----------------------------------------------------------------------------------------------------------------------------------------------
\subsection{General scaling rule}
\label{app:gsr}
Consider a bipartition on the honeycomb lattice shown in Fig.~\ref{fig:2d0} (a). The parallelogram is expanded by two unit vectors $\vec{n}_1 = (1/2, \sqrt{3}/2)$ and ${\vec{n}_2} = (-1/2, \sqrt{3}/2)$ with a total size $\Omega = l_x \times l_y$ and the subsystems are chosen as $A = B = (l_x/2) \times l_y = {(L/2)} \times L$. For convenience, we adopt new coordinates $\vec{r} = x\vec{n}_1 + y\vec{n}_2: x = 1, 2, \dots, l_x, y = 1, 2, \dots, l_y$. The summation in the bipartite fluctuations can then be re-expressed into
  \begin{gather}
    \mathcal{F}_{\Omega} = \sum_{\vec{r}, \vec{r}' \in \Omega} I(\vec{r}' - \vec{r}) = \sum_{x,x'=1}^{l_x} \sum_{y,y'=1}^{l_y} I(x'-x, y'-y).
  \end{gather}
  
To derive general scaling arguments in a `simple' way, we only consider the case where $I(\vec{r})$ is an isotropic function of the distance
    $|\vec{r}| = \sqrt{x^2 + xy + y^2}$.
By analogy to Eq.~(\ref{eq:ds0}), a relation between the double and single sums  can be established
  \begin{align} 
   & \sum_{x,x'=1}^{l_x}  I(x'-x, y)  \notag \\
         &= l_x I(0, y) + \sum_{x=1}^{l_x} (l_x -x) \left[ I(x,  y)+I(-x, y) \right]
    \end{align}
 and the same for $ \sum_{y,y'=1}^{l_y} I( x, y' - y)$.
  Then the bipartite fluctuation function can be grouped into four parts:
  \begin{gather}
   \mathcal{F}_{\Omega}  = l_xl_y I_1 + l_x I_2 + I_y I_3 + I_4, \label{eq:gef}
  \end{gather}
  with 
  \begin{align}
     I_1 &= I(0,0) + 2\sum_{x=1}^{l_x} I(x,0) + 2\sum_{y=1}^{l_y} I(0,y) \notag \\
           &\phantom{=} + 2\sum_{x = 1}^{l_x} \sum_{y = 1}^{l_y} \left[ I(x,y) + I(-x, y)\right], \notag \\
     I_2 &= -2\sum_{y=1}^{l_y} y I(0,y) - 2 \sum_{x=1}^{l_x} \sum_{y=1}^{l_y} y \left[ I(x,y)+I(-x,y)\right], \notag \\
     I_3 &= -2\sum_{x=1}^{l_x} x I(x,0) - 2 \sum_{x=1}^{l_x} \sum_{y=1}^{l_y} x \left[ I(x,y)+I(-x,y)\right], \notag \\
     I_4 &= 2\sum_{x=1}^{l_x} \sum_{y=1}^{l_y} xy \left[ I(x,y) + I(-x,y) \right].
    \label{eq:i4}
   \end{align}
   
The dominant scaling terms in $\mathcal{F}_{\Omega}$ depend on the particular form of $I(r)$. Suppose a general case where $I(r) \propto r^{-\alpha}$ and $l_x$, $l_y$ are of the same order as $L$:\\
  \begin{align}
   I_1 &\sim \mathcal{O}(1)+\mathcal{O}(\frac{1}{L^{\alpha - 1}}) + \mathcal{O}(\frac{1}{L^{\alpha-2}}),\notag \\
   I_2, I_3 &\sim \mathcal{O}(1) +  \mathcal{O}(\frac{1}{L^{\alpha - 2}}) + \mathcal{O}(\frac{1}{L^{\alpha-3}}), \notag \\
   I_4 &\sim  \mathcal{O}(1) + \mathcal{O}{(\frac{1}{L^{\alpha - 4}})},  \label{eq:gsr}
  \end{align}
where $\mathcal{O}({1}/{L^0}) \sim \mathcal{O}(\ln L)$. The leading-order scaling in $\mathcal{F}_{\Omega}$ becomes
   \begin{gather}
     I(r) \propto \frac{1}{r^\alpha}, \quad \mathcal{F}_{\Omega} \propto 
       \begin{cases}
         L^4, \quad \alpha = 0; \\
         L^3, \quad \alpha = 1; \\
         L^2, \quad \alpha \ge 2.
       \end{cases}
   \end{gather}
{When  $I(r) \propto e^{-r/\xi}$, $\alpha \to \infty$, $\mathcal{F}_{\Omega}$ still show the volume law: $\mathcal{F}_{\Omega} \propto L^2 = \mathcal{V}$. Besides, after the subtraction:  $ \left| \mathcal{F}_{A \cup B} - \mathcal{F}_A - \mathcal{F}_B \right|/2$, while the higher order terms $L^4$ and $L^3$ survive in $\mathcal{F}_{AB}$ , the square term $L^2$ always vanishes, which leads to an area law in $\mathcal{F}_{AB} \propto L = \mathcal{A}$. 

To evaluate $\mathcal{F}_{AB}$ more precisely, we are going to study next-leading order terms in $\mathcal{F}_{\Omega}$ case by case.}

%--------------------------------------------------------------------------------------------------------------------------------------------
\subsection{Kitaev {model: the gapless phase}}
\label{sec:hcknb}
For the gapless phase of the Kitaev honeycomb model, {first we consider an isotropic form of the valence bond correlator in Eq.~(\ref{eq:i_func})}
  \begin{gather}
    I(x, y) = \frac{\tilde{c}_1}{(x^2 + xy + y^2)^2},
  \end{gather}
{where ${\tilde{c}_1}$ is a constant for a given $J_z \in [0, 0.50]$ in our convention}.

As $\alpha = 4$, from the general scaling rule in Eq.~(\ref{eq:gsr}) we get $I_i (i=1,2,3) \propto \mathcal{O}(1)$ and $I_4 \propto \mathcal{O}(\ln L)$. In particular, due to the convergence of the lattice summations in $I_{i} (i=1,2,3)$ in Eq.~(\ref{eq:i4}), we can introduce a cutoff $l_x = l_y = \Lambda = 10^3$  to approximate these pre-factors:
  \begin{align}
    I_1 &\simeq I(0,0) + 2\sum_{x=1}^{\Lambda} I(x,0) + 2\sum_{y=1}^{\Lambda} I(0,y) \notag \\
          &\phantom{=====} + 2\sum_{x = 1}^{\Lambda} \sum_{y = 1}^{\Lambda} \left[ I(x,y) + I(-x, y)\right]  \notag \\
          &= I(0) + 7.32 \widetilde{c}_1, \notag \\
    I_2 &= I_3 \simeq   -2\sum_{y=1}^{\Lambda} y I(0,y) - 2 \sum_{x=1}^{\Lambda} \sum_{y=1}^{\Lambda} y \left[ I(x,y)+I(-x,y)\right] \notag \\
         &= -7.68\widetilde{c}_1.
  \end{align}
  Here $I(0)$ denotes the on-site contribution and has the value
    \begin{gather}
      I(0)  = 1 - \langle {Q}_i \rangle^2. 
    \end{gather}  
      $\langle {Q}_i \rangle = \langle \sigma^z_{i_1} \sigma^z_{i_2} \rangle$ represents the integral over the Brillouin  zone: 
      \begin{gather}
        \langle {Q}_i \rangle = (2\pi)^{-2} \int_{-\pi}^{\pi} \int_{-\pi}^{\pi} dk_1 k_2 \cos \theta(k_1, k_2),
      \end{gather}
   with $\cos \theta(k_1, k_2) = a/\sqrt{a^2 + b^2}$ and $f(k) = a + bi = J_x e^{ik_1} + J_y e^{ik_2} + J_z$. Combined with the general expression of $\mathcal{F}_\Omega$ in Eq.~(\ref{eq:gef}), we arrive at
   \begin{align}
   \mathcal{F}_{A} &= \alpha L^2 + \beta L + \mathcal{O}(\ln L), \notag \\
    \mathcal{F}_{AB} &= {\alpha}_\mathcal{F}  L + \mathcal{O}(\ln L),
  \end{align}
  where 
    \begin{align}
      \alpha &= I_1/2 = I(0)/2 + 3.66 \tilde{c}_1, \notag \\
      \beta &= 3I_2/2 = -11.5\tilde{c}_1, \notag \\
      {\alpha}_\mathcal{F}  &= |I_2|/2 =  3.84 \widetilde{c}_1.
     \end{align} 
     
     In the special case $J_x = J_y = J_z = 1/3$, we have 
     \begin{gather}
       \langle Q_i \rangle \simeq 0.525, \quad I(0)/2 \simeq 0.362. \label{eq:osc}
      \end{gather} 
       As indicated by the inset of Fig.~\ref{fig:2d0} (b), the pre-factor of the $L^2$  term in $\mathcal{F}_A$ takes the value $\alpha = 0.391$ close to $I(0)/2$. We conclude that the major contribution to $\mathcal{F}_A$ comes from the on-site interactions.        
%-------------------------------------------------------------------------------------------------------------------------------------------
\subsection{Kitaev model: the gapped phases}
\label{sec:hckb}
In the gapped phases, the bond correlation function decays exponentially and there is less anisotropy observed in Fig.~\ref{fig:bb_r} (b). We can safely start with an isotropic form 
  \begin{gather}
    I(x,y) = c_2e^{-\sqrt{x^2+xy+y^2}/\xi}.
  \end{gather}
  
For $\alpha  = \infty$ in Eq.~(\ref{eq:gsr}), all of the amplitudes $I_i (i=1,2,3,4) \propto \mathcal{O}(1)$. Yet we can still relate them to the powers of the finite correlation length $\xi$. The following assumption is taken
   \begin{gather}
       \xi \sim L^{1/p}, \qquad p \ge 5,
   \end{gather}
  such that when $n < 5$,
    \begin{gather}
      L/ \xi \to \infty, \quad 1/\xi \to 0,  \quad \xi^n < L.
    \end{gather}
Back to  Eq.~(\ref{eq:i4}), we can then replace the summations on the lattice with integrals 
    \begin{align}
        &\sum_{x=1}^{l_x} I(x,0)  = c_2 \xi \int_{1/\xi}^{l_x/\xi}  dx e^{-x} = c_2 \xi \int_{0}^{\infty}  dx e^{-x} = c_2 \xi,\notag \\
        &\sum_{y=1}^{l_y} y I(0,y)  = c_2 \xi^2 \int_0^\infty y e^{-y} = c_2 \xi^2, \notag \\
        & \sum_{x = 1}^{l_x} \sum_{y = 1}^{l_y} \left[ I(x,y) + I(-x, y)\right] = 3.63 c_2 \xi^2 \notag \\
       &= c_2 \xi^2 \int_0^{\infty}  \int_0^{\infty}  dxdy \left( e^{-\sqrt{x^2+xy+y^2}}+e^{-\sqrt{x^2-xy+y^2}} \right), \notag \\                                                                                                
       &\sum_{x=1}^{l_x} \sum_{y=1}^{l_y} y \left[ I(x,y)+I(-x,y)\right]  = 5.33c_2\xi^3  \notag \\
        &= c_2 \xi^3 \int_0^{\infty}  \int_0^{\infty}  dxdy \left( e^{-\sqrt{x^2+xy+y^2}}+e^{-\sqrt{x^2-xy+y^2}} \right)y, \notag \\
       &\sum_{x=1}^{l_x} \sum_{y=1}^{l_y} xy \left[ I(x,y)+I(-x,y)\right]  = 10.4c_2 \xi^4 \notag \\
       &= c_2 \xi^4 \int_0^{\infty}  \int_0^{\infty}  dxdy \left( e^{-\sqrt{x^2+xy+y^2}}+e^{-\sqrt{x^2-xy+y^2}} \right)xy.
    \end{align}
   The pre-factors now read
     \begin{align}
      I_1 &= I(0) + 7.26c_2 \xi^2 + 4c_2\xi, \notag \\
      I_2 &= I_3 = -10.7c_2 \xi^3 - 2c_2\xi^2, \notag \\
      I_4 &= 20.8c_2\xi^4.
     \end{align}
     
Correspondingly, the bipartite fluctuations share the quadratic and  linear forms respectively
   \begin{align}
     \mathcal{F}_A &= \alpha L^2 + \beta L + \mathcal{O}(1),\notag \\
     \mathcal{F}_{AB} &= {\alpha}_\mathcal{F}  L + \mathcal{O}(1),
   \end{align}
   with 
   \begin{align}
     \alpha &= I_1/2 \propto \xi^2,  \notag \\
     \beta &= 3I_2/2 \propto \xi^3, \notag \\
     {\alpha}_\mathcal{F}  &= |I_2|/2 \propto \xi^3.
   \end{align} 
  It  indicates that in gapped phases,  ${\alpha}_\mathcal{F}$ increases at the transition towards the intermediate gapless phase. 

%----------------------------------------------------------------------------------
\section{Entanglement entropy of the Fermion sector}
\label{app:entropy}

In this part, we extract the linear factor ${\alpha}_\mathcal{S}$ in $S_F$ applying the same numerical approach as in Refs.~\cite{peschel2003,
  yao2010}.
The general idea is to map the total system $A \cup B$ on the honeycomb lattice of Fig.~\ref{fig:2d0} (a) onto a torus geometry depicted in Fig.~\ref{fig:torus} (a). Periodic boundary conditions (PBC) {are imposed} along $X$ and $Y$ directions and the boundaries {between subsystems now turn} into two circles. Fig.~\ref{fig:torus} (a) also illustrates one of the two zigzag  boundaries located on $x$-links. For this shape of the boundary, the generic Hamiltonian of subsystem $A$ reads
  \begin{gather}
    \mathcal{H} = \sum_{\langle jk \rangle \in A} i J_a c_j c_k u_{\langle j k \rangle_a} + \sum_{\langle \langle jk \rangle \rangle \in A} i\kappa c_jc_k (-u_{\langle ji \rangle_a} u_{\langle ik \rangle_b}), \label{eq:ha}
  \end{gather}
where $\langle \ \  \rangle$ denotes the nearest-neighbor links and $\langle\langle \ \  \rangle\rangle$ the next-nearest-neighbor links. The gauge choice $u_{\langle jk \rangle_a} = +1, -u_{\langle ji \rangle_a} u_{\langle ik \rangle_b} = +1$ is manifested in the direction of the arrows from site $k$ to site $j$ in Fig.~\ref{fig:torus} (a). 

Now along  the $X$ direction, we go to the momentum space ($k_x = 2m\pi/N_x,  m = 0, 1, \dots, N_x - 1$) and Fourier transform  the Hamiltonian (\ref{eq:ha}) into
    \begin{gather}
      \mathcal{H} = \sum_{k_x} \mathcal{C}^\dagger (k_x) M (k_x) \mathcal{C}(k_x)
    \end{gather}
    in the basis of the matter Majorana fermions
    $\mathcal{C}^\dagger  (k_x)= \left( c_{-k_x, 1}, c_{-k_x,2}\right)$. The matrix takes the form
  \begin{gather}
    M(k_x)= 
    \begin{pmatrix}
      \alpha & is & -\beta & & & \\
      -is^* & -\alpha & ir & \beta & & \\
      -\beta & -ir & \alpha & is & -\beta & \\
      & \beta & -is^* & -\alpha & ir & \cdots \\
      & & -\beta & -ir & \alpha &  \cdots \\
      & & & & \cdots & \cdots
    \end{pmatrix},
  \end{gather}
with matrix elements $r= -J_x$, $s = (J_y + J_z) \cos(k_x/2) - i (J_y - J_z) \sin (k_x/2)$, 
   $\alpha = 2\kappa \sin k_x$, $\beta = 2\kappa \sin (k_x/2)$.
It should be noted that the intrinsic structure of our $2N_y \times 2N_y$ Hamiltonian matrix $M(k_x)$ is distinct from Ref.~\cite{yao2010} where only one type of next-nearest-neighbor couplings $J'$ is included. In our case, all three types of next-nearest-neighbor couplings between the matter Majorana fermions are taken into account, {as} they arise naturally from the effects of the uniform magnetic field~\cite{kitaev2006}. Additionally, to be consistent with the bipartition  we choose for  $\mathcal{F}_{AB}$ in Fig.~\ref{fig:2d0} (a), the boundary position has been switched from $z$-links in Ref.~\cite{yao2010} to $x$-links. 

The diagonalized Hamiltonian reads
   \begin{gather}
     \mathcal{H} = \sum_{n, k_x} \xi_{n} (k_x) \left(\psi_{nk_x}^\dagger \psi_{nk_x} - \frac{1}{2}\right),
   \end{gather}
 where $n = 1, \dots, 2N_y$ and $\psi_{nk_x}$ represents the standard complex fermionic annihilation operators.
For a free fermion system, the energy spectrum $\xi_{n}(k_x)$ can be used to calculate the entanglement entropy~\cite{peschel2003}
  \begin{align}
   \mathcal{S}(k_x) 
     = -\frac{1}{2} \sum_{n, k_x} \left[ \lambda_n \log \lambda_n + (1-\lambda_n) \log (1-\lambda_n) \right] (k_x).
     \end{align}
 Here $\lambda_n(k_x) =(e^{\beta \xi_{n}(k_x)}+1)^{-1}$ denote the eigenvalues of the single-particle correlation function $\langle \psi^\dagger_ {nk_x} \psi_{n'k_x} \rangle$. It follows a  Fermi-Dirac distribution with an inverse temperature $\beta = 1/(k_BT)$. Fig.~\ref{fig:torus} (b) shows the numerical entanglement spectrum $\lambda_n(k_x)$ for the {non-Abelian and Abelian} phases. We observe two gapless branches in the non-Abelian phase. These two modes are responsible for the peaks in entanglement entropy plotted in Fig.~\ref{fig:torus} (c). Both features disappear in the Abelian phase. 

Since $\mathcal{S}_F = \sum_{k_x} \mathcal{S} (k_x)$, through the summation of $\mathcal{S}(k_x)$ in the momentum space followed by a finite-size scaling with respect to $N_x$, we are able to obtain the pre-factor $\alpha_\mathcal{S}$ of the linear term in $\mathcal{S}_F$. 
The results are shown in Fig.~\ref{fig:2d} (d) in the Article. 

\begin{figure}[t]
   \begin{center}
		 \includegraphics[width=.5\linewidth]{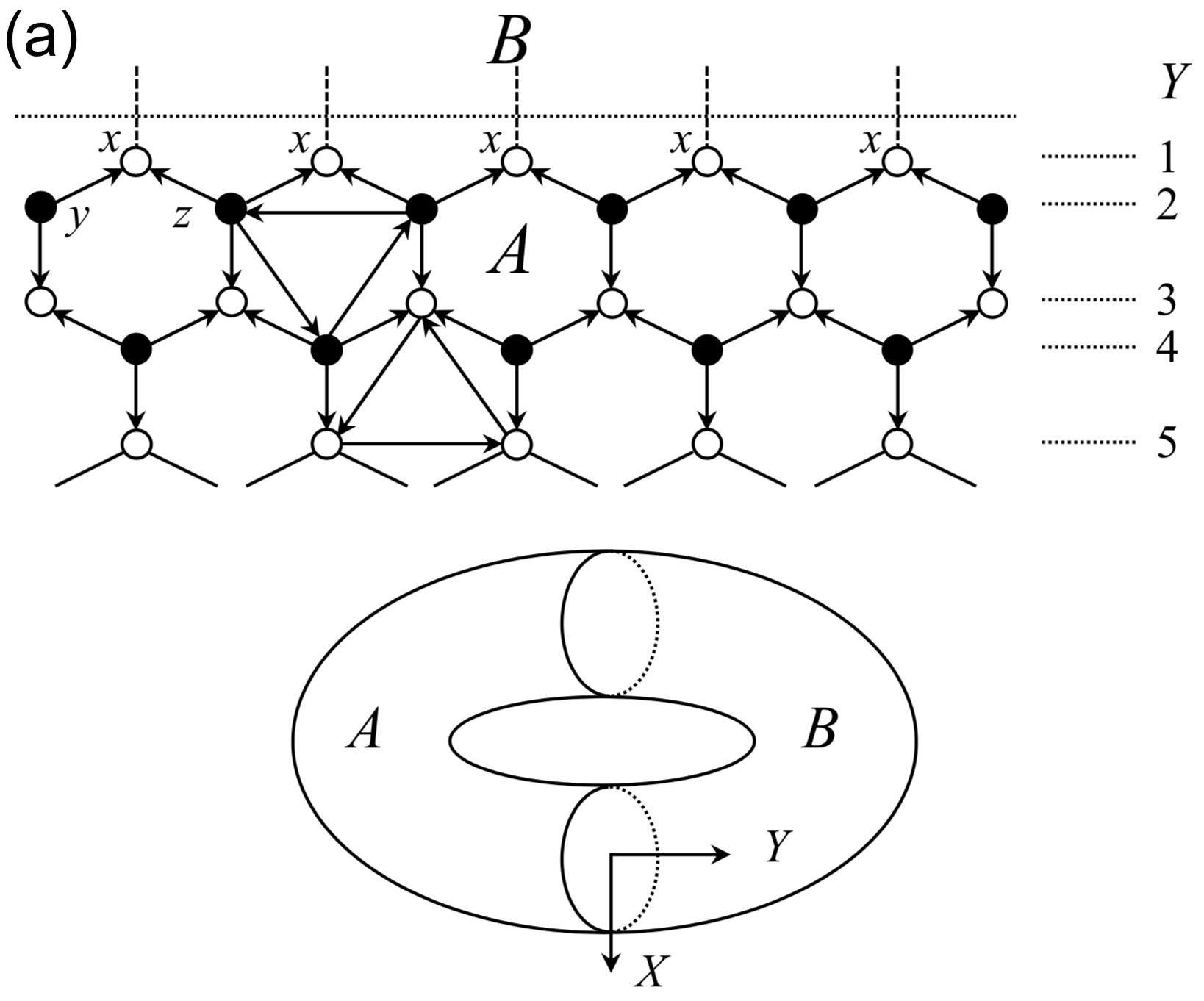}  \\
		 \vspace{0.25cm}
		  \includegraphics[width=.9\linewidth]{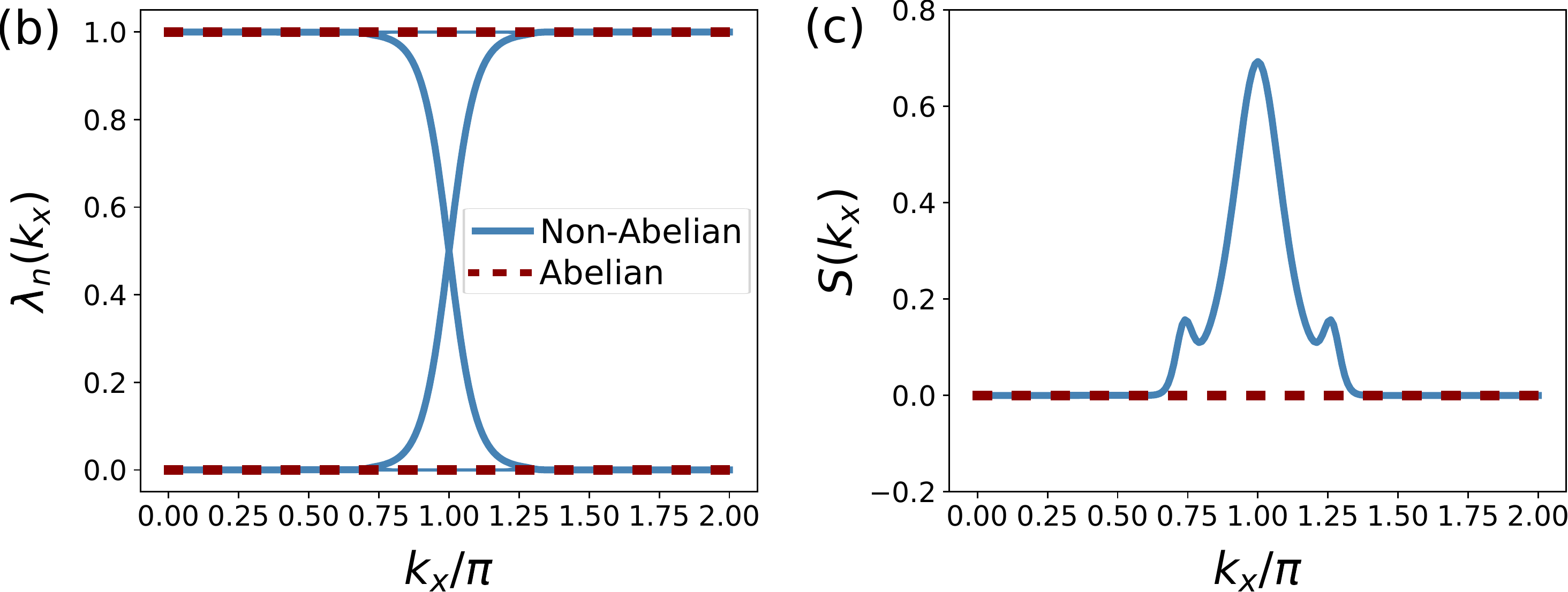}
    \end{center}
		  \vskip -0.5cm \protect\caption[]
     {(a) Bipartite honeycomb lattice mapped onto a torus with two boundary circles along the $X$ direction. Here, we have exchanged the $X$ and $Y$ axes compared to the original lattice in Fig.~\ref{fig:2d0} (a). On the zigzag boundary between subsystems $A$ and $B$, the vertical $x$-links on the first row are suppressed. Two types of  arrows $(j  \longleftarrow k)$ indicate the artificial gauge choice in Hamiltonian (\ref{eq:ha}): $u_{\langle jk \rangle_a} = +1, -u_{\langle ji \rangle_a} u_{\langle ik \rangle_b} = +1$. Entanglement spectrum (b) and entanglement entropy (c) of the non-Abelian and Abelian phases. For the non-Abelian phase (blue lines), we take $\kappa = 0.01$ and $J_x = J_y = 0.30, J_z = 0.40$; for the Abelian phase (red dashed lines), we take $\kappa = 0.00$ and $J_x = J_y = 0.20, J_z = 0.60$. To reflect the low-temperature properties, the inverse temperature $\beta$ is chosen to be $N_x = 100$. The matrix size of $M(k_x)$  is  set as $2N_y \times 2N_y$ {with} $N_y= 100$.}
	\label{fig:torus}
	\vskip -0.5cm
\end{figure}

%---------------------------------------------------------------------------------------
\section{Heisenberg antiferromagnet on honeycomb lattice}
\label{app:sw}
First, we give a review of the modified spin-wave theory on a two-dimensional Heisenberg antiferromagnet.  We then analyze the asymptotic behavior of the two-spin correlation function on the honeycomb lattice. Later, a closed form of the valence bond correlator is derived  and the $r^{-1}$ scaling is verified both analytically and numerically.

%-------------------------------------------------------------------------------------------------------------------
\subsection{Modified spin-wave theory}
\label{sec:hei_sw}
For an antiferromagnetic Heisenberg model on the honeycomb lattice, we can reach the N{\'e}el state by applying a staggered magnetic field~\cite{takahashi1989, song2011}. 

The Hamiltonian reads
  \begin{gather}
    \hat{H} = \frac{J}{2}\sum_{\vec{r},\vec{\delta}} \hat{S}_{\vec{r}} \cdot \hat{S}_{\vec{r}+\vec{\delta}} - h \sum_{\vec{r}} (-1)^{|\vec{r}|} \hat{S}_{\vec{r}}^z.
    \label{eq:h_h}
  \end{gather}
Two sets of sublattices $\{1, 2\}$ are differentiated by $(-1)^{|\vec{r}|} = 1$ for $\vec{r} \in 1$ and $(-1)^{|\vec{r}|} = -1$ for $\vec{r} \in 2$. Vectors $\vec{\delta}$ connect each site to its nearest neighbors, with the total number of nearest neighbor sites denoted by $z$. The introduction of the staggered field breaks the $O(3)$ spin-rotational symmetry and helps to repair the divergence in Green functions arising from the zero mode (or the Goldstone mode).

 In the modified spin-wave theory~\cite{takahashi1989}, one can map spin operators to bosonic operators:
     $ \vec{r} \in 1$, $S_{r}^+ = \sqrt{2S - a_r^\dagger a_r}  \ a_r$, $S_{r}^- = a_r^\dagger \sqrt{2S - a_r^\dagger a_r}$, $S_r^z = S - a_r^\dagger a_r$;
      $\vec{r} \in 2$, $S_{r}^+ = b_r^\dagger \sqrt{2S - b_r^\dagger b_r}$, $S_{r}^- =  \sqrt{2S - b_r^\dagger b_r} \ b_r$, $S_r^z = -\left(S - b_r^\dagger b_r\right)$ with $[a_r, a_{r'}^\dagger] = [b_r, b_{r'}^\dagger] = \delta_{r,r'}$.
   Expansion around large $S$ then gives the Hamiltonian of the order $\mathcal{O}(S^2)$ and $\mathcal{O}(S)$.
   
Combining Fourier transform and Bogoliubov transformation, it is straightforward to obtain single-particle expectation values
  \begin{gather}
    \langle a^\dagger_{\vec{r}} a_{\vec{r}}\rangle = \langle b^\dagger_{\vec{r}} b_{\vec{r}}\rangle
     = -\frac{1}{2} \delta_{\vec{r},\vec{r}'} + f\left(\vec{r}-\vec{r}' \right), \notag \\
      \langle a^\dagger_{\vec{r}} b^\dagger_{\vec{r}'} \rangle       
     = \langle a_{\vec{r}} b_{\vec{r}'} \rangle = g\left(\vec{r}-\vec{r}' \right),
  \end{gather}
  with
  \begin{gather}
   f\left(\vec{r} \right) = \frac{1}{N}\sum_{\vec{k}} \cos(\vec{k}\cdot\vec{r}) \frac{1}{\sqrt{1-(\eta \gamma_k)^2}}, \notag \\
    g\left(\vec{r} \right) = \frac{1}{N}\sum_{\vec{k}} \cos(\vec{k}\cdot\vec{r}) \frac{(-\eta \gamma_k)}{\sqrt{1-(\eta \gamma_k)^2}}, \label{eq:hei_fg}
  \end{gather} 
and others all vanish. Here, $N$ denotes the total number of lattice sites. $\eta$ and $\gamma_k$ are functions depending on the geometry of the lattice
  \begin{gather}
    \eta = \left(1 + \frac{h}{zJS} \right)^{-1}, \notag \\
    \gamma_k = \frac{1}{z} \sum_{\vec{\delta}} \cos(\vec{k}\cdot\vec{\delta}).
 \end{gather}

Let us take a closer look at  a finite honeycomb lattice.  The geometric function reads
 \begin{gather}
   \gamma_k = \frac{1}{3}\left[ \cos{(k_y/\sqrt{3})}+2\cos{(k_x/2)}\cos(\sqrt{3}k_y/6)\right].
 \end{gather}
When $h \to 0$, $\eta \to 1$, there exists one zero mode  $\vec{k}_0 = (0,0)$ making $f(\vec{r})$ and $g(\vec{r})$ divergent.

Following Ref.~\cite{takahashi1989}, one repairs the divergence by adjusting the strength $h$ of the local staggered magnetic field such that the magnetization becomes zero
  \begin{gather}
    \langle S_r^z \rangle = (-1)^{|\vec{r}|} \left( S - \langle n_r \rangle \right) = 0  \Longleftrightarrow  f(\vec{0}) = S + 1/2. \label{eq:f0}
  \end{gather}
It is noted that only the zero mode is regularized by $h \ne 0$ and the sum over the remaining region can be safely approximated by a finite integral at $h=0$. One arrives at
  \begin{gather}
    f(\vec{0}) = \frac{1}{N}\frac{1}{\sqrt{1-\eta^2}} + \frac{1}{2} \int \frac{d\vec{k}}{(2\pi)^2} \frac{1}{\sqrt{1-\gamma_k^2}} = m_0 + 0.754.
  \end{gather}
Taking  $S = 1/2$ in Eq.~(\ref{eq:f0}), we get
   \begin{gather}
       m_0 = \frac{1}{N}\frac{1}{\sqrt{1-\eta^2}} = 0.246.
   \end{gather}
In the same manner,
   \begin{align}
     g(\vec{0}) &= \frac{1}{N}\frac{(-\eta)}{\sqrt{1-\eta^2}} + \frac{1}{2} \int \frac{d\vec{k}}{(2\pi)^2} \frac{(-\gamma_k)}{\sqrt{1-\gamma_k^2}} \simeq -0.692.
   \end{align}

\subsection{Asymptotic behavior of spin-spin correlation}
Now, we can evaluate the behavior of the two-spin correlation function. From Wick's theorem, it differs between sites
  \begin{align}
   & \langle S^z_{\vec{r}} S^z_{\vec{r}'}\rangle - \langle S^z_{\vec{r}} \rangle \langle S^z_{\vec{r}'}\rangle  \notag \\
     &= 
    \begin{cases}
       1/4, & \vec{r}=\vec{r}';\\
       f^2(\vec{r}-\vec{r}')/3, & \vec{r}\neq \vec{r}' \text{ and } \vec{r}, \vec{r}' \in \text{same sublattice}; \\
      -g^2(\vec{r}-\vec{r}')/3, & \text{otherwise}.  
    \end{cases}
    \label{eq:hei_s}
  \end{align}
To restore the spin-rotational symmetry at zero magnetic field,  we have introduced an extra factor $1/3$ to Eq.~(\ref{eq:hei_s}).

At large distances, an expansion of $f(\vec{r})$ around $\vec{k}_0 = (0,0)$ within radius $\xi$ leads to
  \begin{gather}
    f(\vec{r}) \simeq m_0 + \frac{1}{2}\cdot \frac{\sqrt{3}}{2} \int_0^{\xi} \frac{kdk}{(2\pi)^2} \int_0^{2\pi} d\theta \frac{e^{ikr\cos{\theta}}}{k/\sqrt{3}}. \label{eq:f2}
  \end{gather}
Once we take $r \to \infty$,
  \begin{gather}
    f(\vec{r})  = c_0 + \frac{c_1}{ r},
    \label{eq:fg}
  \end{gather}
with $c_0 = m_0 \simeq 0.246$, $c_1 = 3/(8\pi) \simeq 0.119$.
Numerically, we check in Fig.~\ref{fig:bb_h} (a)  the finite-size scaling of the single-particle expectation function $f(\vec{r})$ from Eq.~(\ref{eq:hei_fg}). 
The coefficients read $c_0 = 0.245$, $c_1 = 0.12$ with a power-law index $\alpha = 1.04 $. They agree well with the asymptotic form.
 
Since the approximation $(-\eta \gamma_k) \simeq -1$ is still valid for $\vec{k} \in \Omega \left( 0, \xi\right)$,  one finds over a long distance
 \begin{gather}
   g(\vec{r}) = -f(\vec{r}).  \label{eq:fgr}
  \end{gather} 
 The two-spin correlation function then reveal a power-law $r^{-1}$ decay with alternating signs on the same and different sublattices.
      \begin{figure}[t]
     \begin{center}
	\includegraphics[width=.95\linewidth]{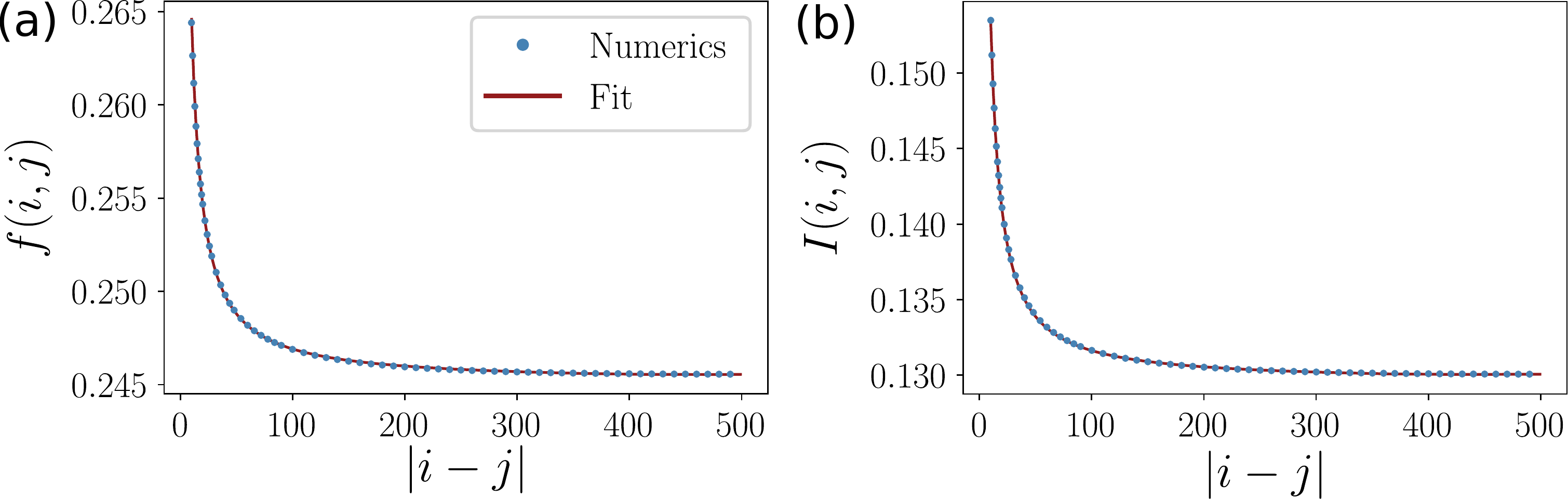}  
     \end{center} 
    \vskip -0.5cm \protect\caption[]
    {Finite-size scaling of correlation functions for the Heisenberg anti-ferromagnetic honeycomb model in the presence of a staggered magnetic field: (a) Function $f(\vec{r})  =  c_0 + c_1 r^{-\alpha}$ with coefficients $c_0 = 0.245$, $c_1 = 0.12$ and $\alpha = 1.04$; (b) Valence bond correlation function $I(\vec{r}) =  c_0 + c_1 r^{-\alpha}$ with coefficients $c_0 = 0.13$, $c_1 = 0.155$ and $\alpha = 1.07$. We set $h = 1.23 \times 10^{-11}$ and the total size $N = 1000$.}
    \label{fig:bb_h}
    \vskip -0.5cm
  \end{figure}     
%---------------------------------------------------------------------------------------
\subsection{Closed form of the valence bond correlator}
\label{sec:hei_bb}
Next, we study the response of valence bond correlation functions in the N{\'e}el state. We adopt the same definition as the Kitaev honeycomb model 
  \begin{gather}
    I (i,j) = \langle Q_i Q_j \rangle - \langle Q_i \rangle \langle Q_j \rangle, \quad Q_i = S_{i_1}^z S_{i_2}^z.
  \end{gather}
 The bond index $\langle i_1 i_2 \rangle$ denotes two sites in the $i$-th unit cell of the sublattice $\{ 1 \}$ and $\{ 2 \}$. In the modified spin-wave theory,  $S_{i_1}^z = 1/2 - a_i^\dagger a_i$, $S_{i,2}^z = -(1/2-b_i^\dagger b_i)$.  Reassembling different terms, we reach
  \begin{align}
    &16I(i,j) \notag \\
    &=  \sum_{k,l = 1,2} \left< \hat{n}_k\hat{m}_l\right> + \left[ 2\left( \langle \hat{n}_1 \hat{n}_2 \rangle + \langle \hat{m}_1 \hat{m}_2 \rangle \right) - \left< \hat{n}_1 \hat{n}_2\right> \left< \hat{m}_1 \hat{m}_2 \right> \right]  \notag \\
    &\phantom{=} - \sum_{k=1,2} \left[ \left< \hat{n}_k \hat{m}_1 \hat{m}_2 \right> +  \left< \hat{n}_1 \hat{n}_2 \hat{m}_k \right>   \right] + \left< \hat{n}_1 \hat{n}_2 \hat{m}_1 \hat{m}_2 \right> - 4.  
  \end{align}
To simplify the notation, we have introduced a new set of number operators 
 \begin{gather}
   \hat{n}_1 = 2a_i^\dagger a_i, \quad \hat{n}_2 = 2b_i^\dagger b_i, \notag \\
   \hat{m}_1 = 2a_j^\dagger a_j, \quad \hat{m}_2 = 2b_j^\dagger b_j. 
 \end{gather}
 
 First, taking into account $\langle a_i^\dagger a_i \rangle = \langle  b_i^\dagger b_i \rangle = - {1}/{2} + f (\vec{0}) = {1}/{2}$, the single-particle expectation values become 
 \begin{gather}
   \langle \hat{n}_1 \rangle = \langle \hat{n}_2 \rangle = \langle \hat{m}_1 \rangle = \langle \hat{m}_2 \rangle = 1.
 \end{gather}. 
 
 Then, Wick's theorem can be applied to calculate the remaining terms involving (2,3,4)-particle expectation values. Before proceeding, from Eq.~(\ref{eq:hei_fg}) we identify three useful functions
  \begin{align}
   i \ne j, & \notag \\
    &f = f(|\vec{r}_i - \vec{r}_j|) = \langle a_i^\dagger a_j \rangle = \langle b_i^\dagger b_j \rangle = \langle a_j^\dagger a_i \rangle = \langle b_j^\dagger b_i \rangle; \notag \\
                        &g = g(|\vec{r}_i - \vec{r}_j|) = \langle a_i^\dagger b_j^\dagger \rangle = \langle a_i b_j  \rangle = \langle a_j^\dagger b_i^\dagger \rangle = \langle a_j b_i \rangle;  \notag \\
   i = j, & \notag \\
    & g_0 =  \langle a_i^\dagger b_i^\dagger \rangle = \langle  a_i  b_i \rangle = -0.692. 
   \end{align}
 All the other terms like $\langle a_i^\dagger a_j^\dagger \rangle$ and $\langle a_i^\dagger b_j \rangle$ vanish in the ground state. Two-particle expectation values then  read
  \begin{align}
    \langle \hat{n}_1 \hat{m}_1 \rangle & = 1 + 4f^2 = \langle \hat{n}_2 \hat{m}_2 \rangle,\notag \\
      \langle \hat{n}_1 \hat{m}_2 \rangle &= 1 + 4g^2 = \langle \hat{n}_2 \hat{m}_1 \rangle, \notag \\
      \langle \hat{n}_1 \hat{n}_2 \rangle &= 1 + 4g_0^2 = \langle \hat{m}_1 \hat{m}_2 \rangle.
   \end{align}
The three-particle expectation value takes the form
  \begin{gather}
    \langle \hat{n}_1\hat{m}_1\hat{m}_2\rangle 
    = 1 + 4g_0^2 + 4(f^2 + g^2) + 16g_0 fg.
  \end{gather}
 The summation is invariant under the exchange of $(i \leftrightarrow j)$. Combined with the sublattice symmetries, we have
  \begin{gather}
    \langle \hat{n}_1\hat{m}_1\hat{m}_2\rangle = \langle \hat{n}_1\hat{n}_2 \hat{m}_1\rangle 
    = \langle \hat{n}_1\hat{n}_2\hat{m}_2\rangle = \langle \hat{n}_2\hat{m}_1\hat{m}_2\rangle.
  \end{gather}
  For the four-particle expectation value, we verify
    \begin{gather}
       \langle \hat{n}_1 \hat{n}_2 \hat{m}_1 \hat{m}_2 \rangle 
       =  \left[(1+4g_0^2) + 4(f^2+g^2)\right]^2+ 64g_0 fg.
  \end{gather}

Therefore, we arrive at a closed form of the bond-bond correlators 
    \begin{align}
      I(i,j) = 2g_0^2 (f^2 + g^2) + (f^2 + g^2)^2.
    \end{align}
    
Considering the asymptotic behaviors of $f(\vec{r})$ and $g(\vec{r})$ functions in Eqs.~(\ref{eq:fg}) and (\ref{eq:fgr}), we establish
   \begin{gather}
      I(i,j) = c_0 + \frac{c_1}{r} + \mathcal{O}(r^{-2}),
     \label{eq:i_hch}
   \end{gather}
 where 
      $c_0 =  4m_0^2 ( g_0^2 + m_0^2 )\simeq 0.131$ and $c_1 = {m_0 (3g_0^2 + 6m_0^2 )}/{\pi} \simeq 0.141$.
Fig.~\ref{fig:bb_h} (b) confirms numerically the  long-range behavior of the valence bond correlator with a power-law index $\alpha = 1.07$. The two pre-factors $c_0 = 0.13, c_1 = 0.155$ are also consistent with our analytical predictions. 
 
 In the end, we find for the N{\'e}el state supported by strong antiferromagnetic Heisenberg exchanges, that the valence bond correlator
  gives a signature of the $r^{-1}$ scaling 
 accompanied by a non-vanishing constant from finite-size effects (the regularization of the zero mode). This is clearly distinct from the pure $r^{-4}$ scaling in the gapless Kitaev spin liquid phase.

%---------------------------------------------------------------------------------------------------------------------------------------------------

\bibliographystyle{apsrev4-1}
\bibliography{sample}

%---------------------------------------------------------------------------------------------------------------------------------------------------

\end{document}